\documentclass[fleqn,usenatbib]{mnras}
\usepackage{graphicx}			
\usepackage{amsmath,amssymb}
\usepackage[caption=false]{subfig}
\usepackage{soul}
\usepackage{booktabs}
\usepackage{scrextend}
\usepackage{xcolor}
\usepackage[flushleft]{threeparttable}

\DeclareUnicodeCharacter{2212}{-}

\newcommand{\msun}{M_\odot}
\newcommand{\mzams}{M_{\rm ZAMS}}
\newcommand{\minmass}{M_\textrm{min}}
\newcommand{\maxmass}{M_\textrm{max}}
\newcommand{\tmin}{t_\textrm{min}}
\newcommand{\tmax}{t_\textrm{max}}

\newcommand{\historicminmassvalue}{8.60^{+0.37}_{-0.41}}

\newcommand{\historicalphavalue}{-2.61^{+1.05}_{-1.18}}
\newcommand{\historicminage}{5.62}
\newcommand{\historiclogminage}{6.75}
\newcommand{\historictmaxvalue}{35.1^{+4.1}_{-3.0}}
\newcommand{\historicbetavalue}{0.13^{+0.83}_{-0.74}}

\newcommand{\match}{{\fontfamily{qcr}\selectfont Match}}
\newcommand{\hybridMC}{{\fontfamily{qcr}\selectfont hybridMC}}

\newcommand{\nsn}{22}

\defcitealias{diaz-rodriguez2018}{DR18}
\defcitealias{williams2014}{W14}
\defcitealias{williams2018}{W18}

\title[Progenitor Mass Distribution of Historic CCSNe]{Progenitor Mass Distribution for $\nsn$ Historic Core-Collapse Supernovae}

\author[M. D\'iaz-Rodr\'iguez et al.]{Mariangelly D\'iaz-Rodr\'iguez,$^{1}$\thanks{E-mail: md14u@my.fsu.edu}
	Jeremiah W. Murphy,$^{1}$\thanks{E-mail: jwmurphy@fsu.edu}
	Benjamin F. Williams,$^{2}$
	Julianne J. 
	\newauthor Dalcanton$^{2}$ and Andrew E. Dolphin$^{3,4}$
	\\
	$^{1}$Department of Physics, Florida State University, Tallahassee, FL 32304, USA\\
	$^{2}$Department of Astronomy, Box 351580, University of Washington, Seattle, WA  98195, USA\\
	$^{3}$Steward Observatory, University of Arizona, Tucson, AZ 85719, USA\\
	$^{4}$Raytheon, Tucson, AZ 85734, USA
}

\date{Accepted 2021 June 13. Received 2021 May 24; in original form 2021 January 22 }

\pubyear{202}

\begin{document}
\label{firstpage}
\pagerange{\pageref{firstpage}--\pageref{lastpage}}
\maketitle
	
	
	\begin{abstract}
		We infer the progenitor mass distribution for $\nsn$ historic core-collapse supernovae (CCSNe) using a Bayesian hierarchical model. For this inference, we use the local star formation histories to estimate the age for each supernova (SN). These star formation histories often show multiple bursts of
		star formation; our model assumes that one burst is associated
		with the SN progenitor and the others are random bursts of star formation. The
		primary inference is the progenitor age distribution. Due to the
		limited number of historic SNe and highly uncertain star formation at young ages, we restrict our inference to the slope of the age distribution and the maximum age for CCSNe. Using single-star evolutionary models, we transform the progenitor age distribution into a progenitor mass distribution.  Under these assumptions, the minimum mass for CCSNe is $\minmass~=~\historicminmassvalue \ \msun$ and the slope of the progenitor mass distribution is $\alpha = \historicalphavalue$. The power-law slope for the progenitor mass distribution is consistent with the standard Salpeter initial mass function ($\alpha = -2.35$). These values are consistent with previous estimates using precursor imaging and the age-dating technique, further confirming that using stellar populations around SN and supernova remnants is a reliable way to infer the progenitor masses.
		
	\end{abstract}
	
	\begin{keywords}
		stars: massive -- supernovae: general -- methods: statistical
	\end{keywords}


\section{Introduction}
\label{sec:introduction}

Stellar evolution predicts that single massive stars with initial masses
$\gtrsim$ 8-11 $\msun$ likely end their lives with core collapse, and a
large fraction of them likely explode as a {\em core-collapse
	supernova} (CCSN;
\citealt{woosley2002,eldridgetout2004,smartt2009}). While there is
some theoretical progress in understanding which stars explode and
which collapse to form black holes, predicting which stars
actually explode remains uncertain
\citep{burrows93,murphy2008,murphy2017,mabanta2018,ugliano2012,sukhbold2016,ebinger2018}. Hence,
it is important to constrain the theory by using observations to
infer the progenitor mass distribution for these explosions.

\citet[][hereafter
\citetalias{diaz-rodriguez2018}]{diaz-rodriguez2018} used the stellar populations associated with 94 supernova remnants (SNRs) to infer the progenitor mass distribution.
\citetalias{diaz-rodriguez2018} found that the minimum mass ($\minmass$) for CCSNe is $7.33^{+0.02}_{-0.16} \ \msun$, the maximum mass ($\maxmass$) for explosion is greater than $59~\msun$, and the power-law slope in between is $\alpha = -2.96^{+0.45}_{-0.25}$; these represent the tightest constraints on the progenitor mass
distribution to date. The analysis assumed that SNRs are unbiased tracers of recent supernovae (SNe). However, SNe and SNRs represent two different tracers of SN location and using SNRs as SN tracers may lead to a biased sample. As a preliminary test of this bias, we infer the progenitor age distribution and progenitor mass distribution of $\nsn$ historic SNe. 

Generally, there are two basic techniques to infer SNe progenitor
masses. The first involves the direct identification of the
precursor that led to an SN (e.g. \citealt{smartt2002em,smartt2002ap,Smartt2004,smarttetal2009,vandyk2003du,vandyk2003gd,maund2004,vandyk2011,vandyk2012a,vandyk2012rsg,li2005,li2006,li2007,maundsn2005cs,maund2011,maund2014bk,hendry2006,galyam2009,smartt2009,smith2011,fraser2014}); the other is age-dating the stellar populations associated with the
SN. Estimating the initial mass of an SN precursor requires
inferring the bolometric luminosity and temperature of the precursor
and comparing to stellar evolution models to infer the progenitor
mass (see \citealt{vandyk2017}). While it is
important to image the star that actually exploded, the direct imaging technique relies on
the last, most uncertain stages of stellar evolution \citep{farrell2020}.

Directly imaging the precursor has led to new constraints on the progenitors of CCSNe.
One of the earliest progenitor studies found a minimum mass of
$\minmass=8.5^{+1.0}_{-1.5} \msun$ and a maximum mass of $\maxmass =
16.5 \pm 1.5 \msun$ for 8 direct Type IIP SN progenitor
detections and 12 upper limits
\citep{smarttetal2009}, assuming a progenitor
distribution that matches the slope of the Salpeter initial mass
function ($\alpha=-2.35$). As expected, all
directly imaged progenitors of SN IIP are red supergiants (RSGs). The luminosity limit for observed RSGs in the local group is $\mathrm{log}(L/L_\odot) = 5.5$, and for single-star models
this corresponds to an initial progenitor mass of $\sim 30 \ \msun$. The lack of detected high mass progenitors above $\sim 17 \msun$ (or luminosities above $\mathrm{log}(L/L_\odot) > 5.2$) in the progenitor sample in \citet{smarttetal2009} suggested that the most massive
RSGs may not be exploding as SN IIP. Later, in an updated and extended sample, \citet{smartt2015} again found that there were no RSGs above $\sim~17 \ \msun$. 

Since this original suggestion of a red supergiant problem, a
debate has developed about the statistical robustness of this
result (see
\citealt{Walmswell2012,daviesbeasor2018,beasor2016}). For example,
\citet{daviesbeasor2018} revisited the bolometric correction and other sources of systematics in
\citet{smarttetal2009} and \citet{smartt2015}. They found a higher upper mass
limit of $\maxmass = 19.0^{+2.5}_{-1.3} \ \msun$, with a 95~\%
confidence limit of $<27 \ \msun$. Based on their study, they
concluded that the evidence for a population of `missing' stars had
only a minor statistical significance. More recently,
\citet{Kochanek2020} used a Bayesian approach to re-analyze the sample
used by \citet{smartt2015} and \citet{daviesbeasor2018}, finding a
maximum mass of $\maxmass = 19.01^{+4.04}_{-2.04} \ \msun$. This study used a one-sigma discrepancy to conclude that the red supergiant problem remains. However, \citet{colquhoun2014} conclude that in practical studies, uncertainties are often underestimated and that even a two-sigma discrepancy is wrong 30\% of the time. They show that in practice, one needs to use the three-sigma rule to keep the false discovery rate under 5\% \citep{colquhoun2014}. Hence, the conclusion that red supergiant problem remains is premature given that the
\citet{Kochanek2020} result is less than
two sigma away from the expected upper limit of 25-30 $\msun$ for RSGs \citep{smarttetal2009}. In fact, these analyses are complicated by the many assumptions in the likelihood models selected; given these assumptions, it is easily possible for the uncertainties to be much larger than inferred.

More recently, \citet{DaviesandBeasor2020} propose a simpler
statistical test.  Given the observed RSG brightness distribution and that one is drawing a small sample, they calculate the probability of the brightest observed RSG
being $\log(L/L_{\odot}) = 5.24 \pm 0.08$ (progenitor of SN S009hd) 
given that the true theoretical brightness limit is $\log(L/L_{\odot} = 5.5)$.  They found that given the small
number of observed progenitors to date, the current brightest SN IIP
progenitor is consistent (with 95\% confidence) with the observed RSG
brightness distribution.

Considering the difficulty in obtaining large
numbers of observed progenitors, there is a clear need for an alternative
technique for inferring the progenitor mass distribution.
While SN progenitors are usually visible as resolved stars in
high-resolution images (\citealt{smartt2009,smartt2015} and references
therein), determining which progenitor
explodes successfully remains 
uncertain because only a few progenitors have been discovered directly
\citep{leonard2010}. This is either due to the discovery rate of SN progenitor stars in nearby galaxies or due to instrument limitations (e.g. depth, field of view). Currently, there are approximately only 30 direct detections and 38 upper limits,  making the statistical constraints quite loose \citep{vandyk2017}. 

An alternative technique is to age-date the stellar population
associated with the SN, and from this age, infer a mass.  These
age-dating techniques offer a way to increase the
number of progenitor mass estimates \citep{maizapellaniz2004, wang2005,crockett2008, gogarten2009,vinko2009,badenes2009,murphy2011,murphy2018,williams2014,williams2018,maund2017,maund2018,Auchettl2019}. This second technique to infer progenitor masses has the advantage in that it does not need a precursor image of the star before it exploded. Instead, it relies on age-dating the stellar populations in the vicinity of the star that exploded, which can be performed after the explosion.

The age-dating technique has been used to infer the progenitor masses of SNe \citep{panagia2000,murphy2011,williams2014,williams2018} and SNRs
\citep{badenes2009,jennings2012,jennings2014,diaz-rodriguez2018,Auchettl2019}. Previous studies
includes deriving the progenitor mass for specific SNe (see
\citealt{murphy2011, williams2014, murphy2018}), whereas other studies used bigger samples to derive the overall progenitor mass distribution using the age-dating technique. For
example, \citet{jennings2012} studied 53 SNRs in M31 and found a
$\minmass$ between $7.0 - 7.8 \ \msun$ for a $\maxmass$ of $\sim 26
\ \msun$ and standard Salpeter IMF slope of $\alpha = -2.35$. In a
later study, \citet{jennings2014} used 115 SNRs in M31 and M33, and
derived a progenitor mass distribution for CCSNe with a maximum mass
of $\maxmass=35^{+5.0}_{-4.0} \ \msun$ and slope $\alpha = -2.35$;
while assuming a minimum mass of $\minmass = 7.3 \ \msun$ as found previously
in \citet{jennings2012}. On the other hand, they found a slope of
$\alpha = 4.2^{+0.3}_{-0.3}$ when fixing the $\maxmass$ to $90~\msun$. \citetalias{diaz-rodriguez2018} used Bayesian
inference to infer all three parameters simultaneously using the SFHs for M33 
\citep{jennings2014} and the SFHs for M31
\citep{lewis2015}.  \citetalias{diaz-rodriguez2018}
found that the $\minmass = 7.33^{+0.02}_{-0.16} \ \msun$, the $\maxmass > 59 \ \msun$, and the power-law slope in between is $\alpha=-2.96^{+0.45}_{-0.25}$ for 94 SNRs in M31 and
M33. Using a similar age-dating technique, \citet{Auchettl2019} found a mass
distribution that is consistent with a Salpeter IMF distribution using
the stellar populations around 23 known SNRs in the Small Magellanic
Cloud. Together, these age-dating techniques have increased the number of progenitor mass estimates by roughly a factor of ten.

While age-dating SNRs allows for a larger sample of progenitor mass
estimates, there remains the possibility that the SNR catalogs are
biased tracers of SNe.  For example, \citetalias{diaz-rodriguez2018}
found that the power-law slope for the progenitor mass is $\alpha=-2.96^{+0.45}_{-0.25}$, which
is steeper than the standard Salpeter IMF.  There are two interpretations
of this steeper slope. Either the most massive stars are exploding less frequently, or there is a bias against finding SNRs in the youngest
star formation (SF) regions. Comparing the progenitor mass distribution
for SNe and SNRs would help to address whether SNR catalogs are biased.

In this paper, we infer the progenitor age distribution using the star formation histories (SFHs) around $\nsn$ historic SNe. We use a Bayesian inference framework previously derived in \citetalias{diaz-rodriguez2018} to infer the parameters of the distribution. In contrast to our previous study \citepalias{diaz-rodriguez2018}, we only infer the maximum age ($\tmax$) and slope of the age distribution ($\beta$), while fixing the minimum age ($\tmin$) to $\historicminage$~Myr. We then use stellar evolution models \citep{marigo2017} to translate the progenitor age distribution parameters to its counterpart in mass space. 

An outline of the paper is as follows. In
Section~\ref{subsection:snsample}, we describe the SN sample used in
this study. We then provide a summary of our Bayesian inference
technique to infer the CCSN progenitor age distribution in
sections~\ref{subsection:bayesianinference} and \ref{subsection:agedistribution}. In section~\ref{section:results},
we present the statistical results. In
section~\ref{section:discussionandconclusion}, presents the conclusions and a discussion of our results in the context of previous progenitor analyses.


\section{Methods}
\label{section:methods}

\begin{table*}
	\begin{threeparttable}
		\caption{SN sample$^a$ used for this analysis.}
		\label{tab:historicsnetable}
		\begin{tabular}{clcclccc}
			\toprule
			SN & Type & {R.A.} & {Dec.} & {Galaxy} & {Mpc}  & {$N_\mathrm{stars}$} & {Reference}\\
			\midrule
			SN1917A & II &308.69542 & 60.12472 & NGC6946  & 6.8 & 52  & W18    \\
			SN1923A & II P & 204.28833 & -29.85111 & NGC 5236 (M83)  & 5.0 & 142  & W14    \\
			SN1948B & II P & 308.83958 & 60.17111 & NGC 6946 & 6.8 & 201 & W18\\
			SN1951H & II & 210.98042 & 54.36139 & NGC 5457 (M101) & 7.2 & 11 & W14 \\
			SN1954A & Ib & 183.9445833 & 36.2630556 & NGC 4214 & 3.0 & 31 & W18\\
			SN1968D & II & 308.74333 & 60.15956 & NGC 6946 & 6.8 & 86 & W18\\
			SN1978K & II & 49.41083 & −66.55128 & NGC 1313 & 4.1 & 19 & W18\\
			SN1980K & II L & 308.875292& 60.106597 & NGC 6946 & 6.8 & 38 & W18\\
			SN1985F & Ib &190.38754 & 41.15164 & NGC 4618 & 7.9 & 203 & W18\\
			SN1987A & II pec & 83.86675 & -69.26974 & LMC & 0.05 & 11800 & W14  \\
			SN1993J & IIb & 148.85323 & 69.02047 & NGC 3031 (M81) & 4.0 & 143 & W14 \\
			SN1994I & Ic & 202.47530 & 47.19181 & NGC5194 & 8.3 & 42 & W14 \\
			SN2002ap & Ic pec &24.099375 & 15.753667& NGC 0628 & 10.0 & 5 & W18 \\
			SN2002hh & II P & 308.68454 & 60.12194 & NGC 6946 & 6.0 & 66 & W14 \\
			SN2003gd & II P & 24.177708 & 15.738611 & NGC 0628 & 10.0 & 20 & W18\\
			SN2004am & II P & 148.94421 & 69.67725 & NGC 3034 (M82) & 4.0 & 37 & W14 \\
			SN2004dj & II P & 114.32092 & 65.59939 & NGC 2403 & 4.0 & 127 & W14 \\
			SN2004et & II P & 308.85554 & 60.12158 & NGC 6946 & 6.8 & 18  & W18\\
			SN2005cs & II P & 202.47237 & 47.17450 & NGC 5194 & 8.3 & 33 & W14 \\
			SN2008bk & II P & 359.46008 & -32.55597 & NGC 7793 & 4.0 & 248 & W14\\
			SN2008iz & II & 148.96479 & 69.67939 & NGC 3034 (M82) & 4.0 & 56 & W14 \\
			SN2017eaw & II P &308.684333 & 60.193306 & NGC 6946 & 6.8 & 136 & W18\\	
			\bottomrule
		\end{tabular}
		\begin{tablenotes}
			\small
			\item {\textbf{Note.$^a$} Columns are (1) name of SN, (2) SN spectral type, (3) right ascension of the SN, (4) declination of the SN, (5) host galaxy of the SN, (6) distance to the galaxy in Mpc, (7) the number of stars in the photometry within the physical radius of 50 pc, and (8) article reference. For more information about the SN sample see \citetalias{williams2018}.}
		\end{tablenotes}
	\end{threeparttable}
\end{table*}

\subsection{SN Sample}
\label{subsection:snsample}
We infer the age distribution for a total of
$\nsn$ historic CCSNe within $\sim$8 Mpc that also have high quality
overlapping HST imaging  (see Table~\ref{tab:historicsnetable}). 11 of
the supernovae (SNe) presented in this manuscript were published initially by
\citet[][hereafter \citetalias{williams2014}]{williams2014} and the
other 11 SNe are from an extended and updated study in
\citet[][hereafter \citetalias{williams2018}]{williams2018}. These historical CCSNe have of order arcsecond accuracy in their positions and modest foreground extinction. The authors used the {\it HST} archive to determine which CCSNe have imaging in two or more broadband filters in ACS, WFPC2 (WFC3 for the most recent SNe sample), or UVIS to cover as much region as possible of the host galaxies.

In Table~\ref{tab:historicsnetable}, we include the SN names considered in this study, SN spectral
type, host galaxy, number of stars in the photometry,
etc. \citetalias{williams2014} and \citetalias{williams2018} used a
combination of HSTphot and DOLPHOT (updated HSTphot
\citep{dolphin2000,dolphin2016}) to obtain the photometry for each
SN. See \citetalias{williams2014} and
\citetalias{williams2018} for more information about exposure dates, durations, and depths for the SN sample.

\subsection{Star Formation Histories}
To infer the progenitor age distribution for these $\nsn$ SNe, we must estimate the age of the stellar populations near each SN.  We use the SFH for the surrounding stellar population as this age estimate. The technique used to obtain the SFH for the sample
in this study has been described and tested in several other
studies. For example, \citet{murphy2011} showed that applying this
technique to SN2011dh is consistent with direct-imaging estimates,
\citet{jennings2014} derived the SFHs for 115 SNRs in M31 and M33,
\citet{gogarten2009} inferred the age and mass for an unusual
transient in NGC 300, \citetalias{williams2018} inferred the progenitor masses for 25 Historic SNe, \citet{murphy2018} estimated the age and mass for the black hole candidate N6946-BH1 and \citet{williams2019} inferred the SFHs for 237 optically identified SNRs in M83.

\begin{figure*}
	\centering
	\includegraphics[width=0.8\textwidth]{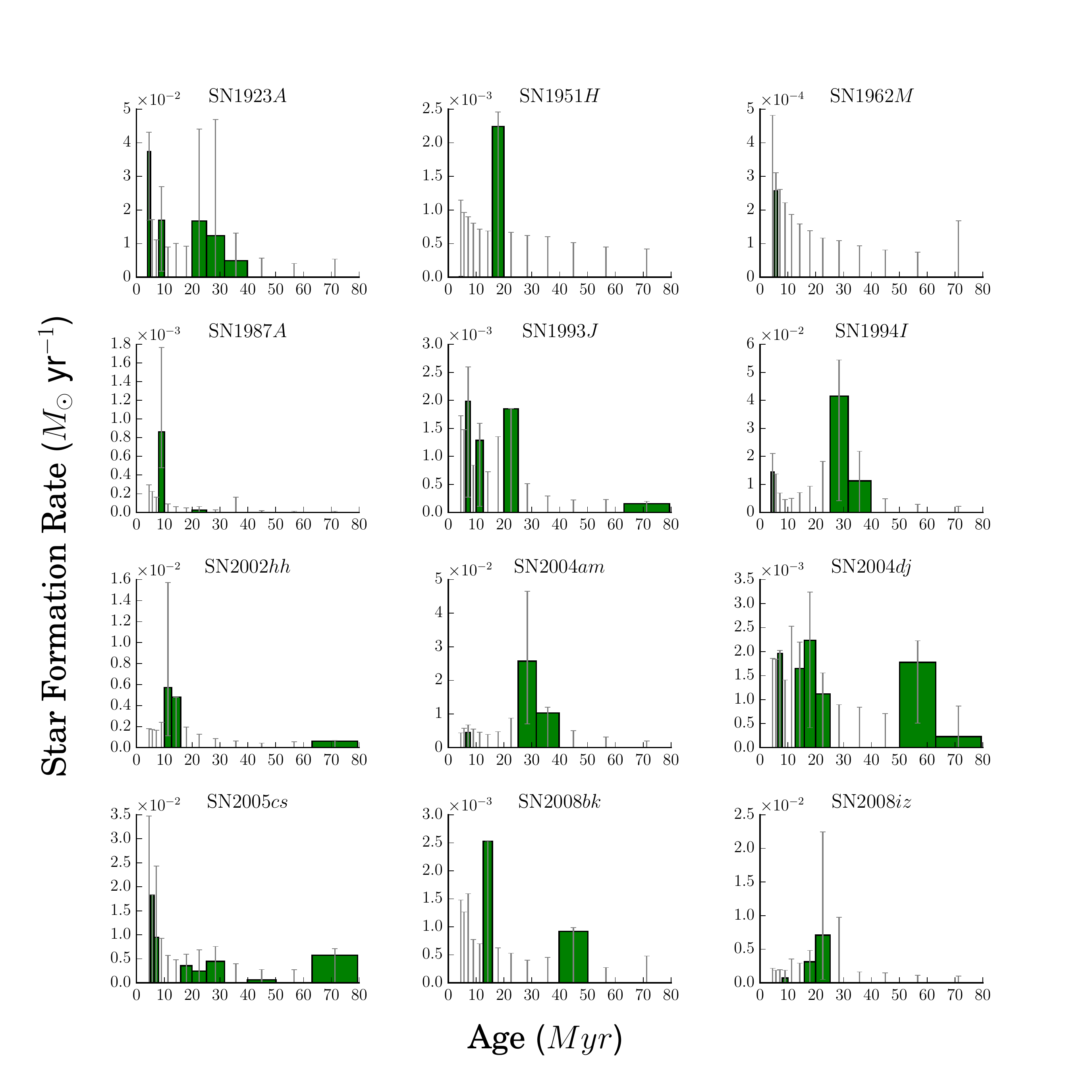}
	\caption{Differential star formation history (SFH) for 50-pc radius regions surrounding historic supernovae (SNe). The bars show the most likely SFH for each age bin, and the uncertainties show a 68\% confidence interval from the hybrid-MCMC posterior. These 12 historical SNe are from \citetalias{williams2014}. The easiest cases to interpret are when the SFH identifies one clear age (e.g. SN1951H, SN1962M, and SN1987A). In these cases, one may associate the age with the progenitor that exploded, and from that age, one may derive the corresponding mass. When there are multiple bursts of SF, it is not clear which burst is associated with the explosion. For these cases, we assume that only one burst is associated with the progenitor explosion; the others are random unassociated bursts of SF.}
	\label{fig:w14historicsfhs}
\end{figure*}

\begin{figure*}
	\centering
	\includegraphics[width=0.8\textwidth]{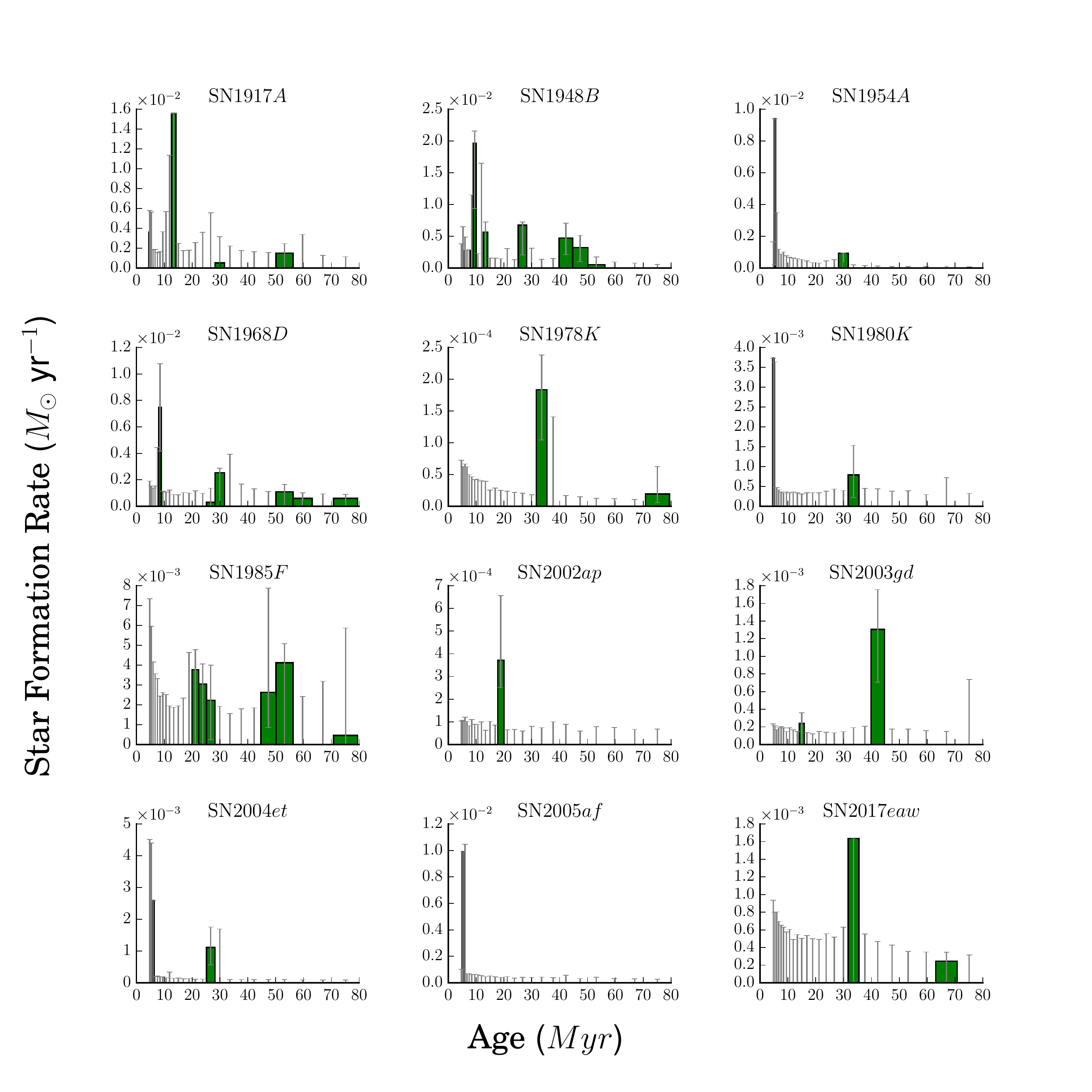}
	\caption{Same as Figure~\ref{fig:w14historicsfhs}. These 12 historical SNe are from \citetalias{williams2018}. }  
	\label{fig:w18historicsfhs}
\end{figure*}

Figures~\ref{fig:w14historicsfhs}~and~\ref{fig:w18historicsfhs}
show the SFHs for the $\nsn$ historic SNe sample used in this
study. \citetalias{williams2018} inferred the SFH for each SN using {\match} 2.7
\citep{dolphin2002,dolphin2012,dolphin2013}. {\match} infers SFHs from observed color-magnitude diagrams (CMDs). \citetalias{williams2018} used the DOLPHOT package to calculate the photometry of all
resolved stars within 50 pc of each SN. The same photometry package, in combination with HSTphot, was used to obtain the photometry for each SN from \citetalias{williams2014}. Essentially, {\match} derives these SFHs by generating a set of theoretical grids in color and magnitude using stellar evolution models of \citet{girardi2010} and \citet{marigo2017}.

{\match} generates SFHs that most accurately represent the observed CMD of the region while taking into account the effects of observational errors, foreground extinction, and distance. While it is possible to fit a single isochrone to an observed CMD, due to the small numbers of massive stars and poor Poisson sampling of the upper end of the IMF, one can underestimate the turnoff mass \citep{williams2018}. Essentially, {\match} fit superpositions of stellar populations to reproduce the observed CMD \citep{dolphin2002, gogarten2009, jennings2012}. As a result, the estimated recent SFH fits the turnoff and full luminosity function of the main sequence. Fitting the entire CMD and not just the turnoff adds significant statistical weight in age-dating the stellar population. The technique of using this fitting package and models has been tested against direct pre-cursor measurements and has shown to have a good agreement (e.g., \citealt{dolphin2003, gallart2005,williams2014,williams2018}, and many others). The work in this manuscript consists of using these SFHs to draw conclusion about the CCSNe theory, for more details about the CMD fitting see \citetalias{williams2018} and references therein. For a comparison of the 22 historic progenitor mass estimates used in this study, with those inferred using the direct imaging technique, see Figure 2 of \citetalias{williams2018}.

The age-dating technique relies on the fact that the majority of stars are born clustered \citep{lada2003} hence they have a common age ($\Delta t\lesssim 3-5 Myr$) and metallicity \citep{Bovy2016}. This is important, because stars that formed in common events remain physically nearby up to $\sim$~100 pc during the 100 Myr lifetimes. Hence, even after a star has exploded, the remaining stars still provide information about the age population and hence its mass. Previous similar studies has shown that this assumption is appropriate (e.g. \citealt{gogarten2009, murphy2011, jennings2014, badenes2015, sarbadhicary2017, williams2019,Auchettl2019}).

As \citetalias{diaz-rodriguez2018} note, there are three principal sources of uncertainty when inferring the age for each SN. First, the resolution of the SFH affects the accuracy for each age bin. The
resolution for the SNe sample published in \citetalias{williams2014}
is $\Delta \rm{{log}}_{10}($$t/$Myr)~=~0.1 and for the SNe in
\citetalias{williams2018} is $\Delta
\rm{{log}}_{10}($$t/$Myr)~=~0.05. Second, it is common to have multiple bursts of SF associated with the stellar population and we do not know
which burst is due to the actual SNe explosion (see
Figure~\ref{fig:w14historicsfhs} and \ref{fig:w18historicsfhs}). These
multiple bursts usually contribute the most to the uncertainty in inferring the age for each object. Lastly, there is an uncertainty associated with the SF rate for each age bin. In a previous analysis, \citetalias{diaz-rodriguez2018} developed a technique to model the first two uncertainties: uncertainty due to resolution and due to multiple bursts. In this manuscript, we expand upon that work and develop a technique that models all three sources of uncertainties. In particular, we use a re-sampling method (Jackknife) to model the uncertainty in the SFH.

\subsection{Hierarchical Bayesian Inference}
\label{subsection:bayesianinference}
To calculate the age distribution for the $\nsn$ historic SNe,
we use Bayes' theorem to compute the posterior joint probability, $P
(\theta | \textrm{Data})$, where $\theta$ represents the parameters of
the age distribution model. We assume that the SN progenitor age
distribution is a power-law with an index of $\beta$ and a minimum and
maximum age ($\tmin$ and
$\tmax$). The posterior distribution is the probability of the model parameters given the observations:
\begin{equation}
\label{bayestheorem}
P (\theta|{\rm Data}) = \frac{\mathcal{L}({\rm  Data}|\theta) \ P(\theta)}{P({\rm Data})} \, 
\propto \mathcal{L}({\rm  Data}|\theta) \ P(\theta) ,
\end{equation}
where $\mathcal{L}({\rm Data}|\theta)$ is the probability of observing
the data, or also known as the likelihood function. $P(\theta)$ is the prior distribution and $P(\rm Data)$ is the posterior distribution normalization.

The primary source of data are a collection of CMDs of the resolved stars, one for each region surrounding an SN. Hence, our goal is to use Bayes' theorem to find the joint posterior distribution of the model parameters given the set of CMDs, $P(\theta | \{ \textrm{CMD} \})$.

In this manuscript, we present a hierarchical Bayesian inference; the primary nuisance parameters are the SFHs derived by {\match}. For each CMD, {\match} infers the best-fit SFH and it uses a hybrid Monte Carlo (MC) algorithm to estimate a posterior distribution for the SFH, i.e.~$P(\textrm{SFH}_k |
\textrm{CMD}_k)$. Where the index $k$ represents each SN.
In this context, the formal hierarchical posterior distribution
is
\begin{equation}
\label{eq:pjointposteriorinitial}
P(\theta | \{ \textrm{CMD}_k \}) =
 \prod_k \int P(\theta | \textrm{SFH}_k )
P( \textrm{SFH}_k | \textrm{CMD}_k )  d\text{SFH}_k \, .
\end{equation}
 In principle, the basic steps to infer the age
distribution parameters includes estimating $P(\theta | \textrm{SFH}_k)$,
using the posterior distribution given by {\match}, $P(\textrm{SFH}_k | \textrm{CMD}_k)$,
and marginalizing over these results via equation~\ref{eq:pjointposteriorinitial}.

In \citetalias{diaz-rodriguez2018}, P($ \textrm{SFH}_k |
\textrm{CMD}_k $) was essentially a delta function $\delta
  (\mathrm{SFH}_k-\overline{\mathrm{SFH}}_k|\mathrm{CMD}_k)$, where
$\overline{\mathrm{SFH}}_k$ is the best fit SFH. Ideally, one would use
the {\hybridMC} results of {\match} for the distribution of SFH,
$P(\textrm{SFH}_k | \textrm{CMD}_k)$. However, the approximate results
of the {\hybridMC} method and the highly uncertain SFHs presented
problems in using this technique.  For example, we calculated the
median of the best-fit SFH and compared it to the distribution of
medians from the {\hybridMC} distribution.  We found that the
median best-fit was often an outlier in the {\hybridMC} results.  This shows that while the {\hybridMC} results are an estimate for the distribution of SFH, for the low number of stars in our CMDs, they do not adequately represent $P(\textrm{SFH}_k |
\textrm{CMD}_k)$.

Since we do not have an accurate estimate for $P(\textrm{SFH}_k |
\textrm{CMD}_k)$, we use the
data itself and a re-sampling method to estimate the effects of the uncertainties in the SFH. Two common re-sampling methods include the
bootstrap and jackknife methods. The bootstrap method tends to
work better for skewed distributions, while the jackknife tends to
work better for small samples. Since the bootstrap double counts sources, this can affect the accuracy in inferring the posterior
distribution for small samples. Alternatively, the jackknife method involves removing one data point exactly $R$ times, where $R$ is the number of data points, or in this case, the number of SNe.  Therefore, the jackknife avoids the doubling counting problem.

For this analysis, the jackknife re-sampling method is as follows.  The $r$th  estimator for the posterior is
\begin{equation}
\label{eq:posteriorjackknife}
  P_r(\theta|\{\textrm{CMD}_{k^{\prime}}\}) = \int \prod_{k^{\prime}} P(\theta | \textrm{SFH}_{k^{\prime}} )
P( \textrm{SFH}_{k^{\prime}} | \textrm{CMD}_{k^{\prime}} )  d\text{SFH}_{k^{\prime}} \, ,
\end{equation}
where $k^{\prime}$ represents the $k^{\prime}$th SN of a subsample of SNe, and $P( \textrm{SFH}_{k^{\prime}} | \textrm{CMD}_{k^{\prime}}) = \delta( \textrm{SFH}_{k^{\prime}} - \overline{\textrm{SFH}}_{k^{\prime}} | \textrm{CMD}_{k^{\prime}})$.  Each re-sample $r$ has a one less SN than the full
sample, i.e. $N_{k^{\prime}} = N_{\rm SNe} - 1$.  There are $N_{\rm SNe}$ unique
subsamples for jackknife re-sampling, in other words $N_r = N_{\rm SNe}$.  To find the final posterior
distribution, we marginalize over all of the jackknife estimators for
the posterior.
\begin{equation}
 P(\theta|\{\textrm{CMD}_{k}\} \approx \sum_r^{N_{\rm SNe}}
 P_r(\theta|\{\textrm{CMD}_{k^{\prime}}\}) \, .
\end{equation}
The next section discusses the final piece, calculating $P(\theta|\{\textrm{SFH}_k\})$.

\subsection{Inferring the Progenitor Age Distribution}
\label{subsection:agedistribution}
The method to calculate the posterior distribution of $\theta$ given a
set of SFHs, $P(\theta | \{ \textrm{SFH}_{k'} \})$ is described in
\citetalias{diaz-rodriguez2018}. First, we convert each SN SFH into a probability density function (PDF) for the age, $P_{k'} (t)$. {\match} returns a discrete SFH, and the set of age bins is $\{i\}$, where $i$ indexes each age bin. Given this discrete version of the SFH, the probability of a star being associated with age bin $i$ is:
\begin{equation}
\label{eq:probofburst}
P_{\rm SF_{k'}}(i) = P_{k'}(i) \cdot \Delta t_{{k'},i}
= \frac{{\rm SFR_{k'}}(i)}{M_{\star_{k'}}(T_{\rm max})} \cdot \Delta t_{{k'},i} \, .
\end{equation}
This PDF is proportional to the stellar population SFR, the total amount of stars ($M_{\star}$) formed in the last $T_{\rm max}$ Myr, and $\Delta t_i$ represents the width of each age bin $i$. 

To accurately model and estimate the parameter $\tmax$, we include ages above the
predicted canonical value ($\sim 8 \ \msun$, this corresponds to an
age of $\sim 45 \ Myr$) in the single-star scenario. To allow the
algorithm to explore the parameter space fully, we consider ages from $\historicminage$~Myr up to
$T_\mathrm{max}=80$~Myr. Figures~\ref{fig:w14historicsfhs}~and~\ref{fig:w18historicsfhs} show the best-fit SFH with its corresponding uncertainties for our SN sample. In most cases, the best-fit SFH shows more than one burst of SF. Since it is not clear which burst is associated with the progenitor that exploded, one needs to model these unassociated bursts of SFH properly.  

\begin{figure*}
	\centering
	\includegraphics[width=0.7\textwidth]{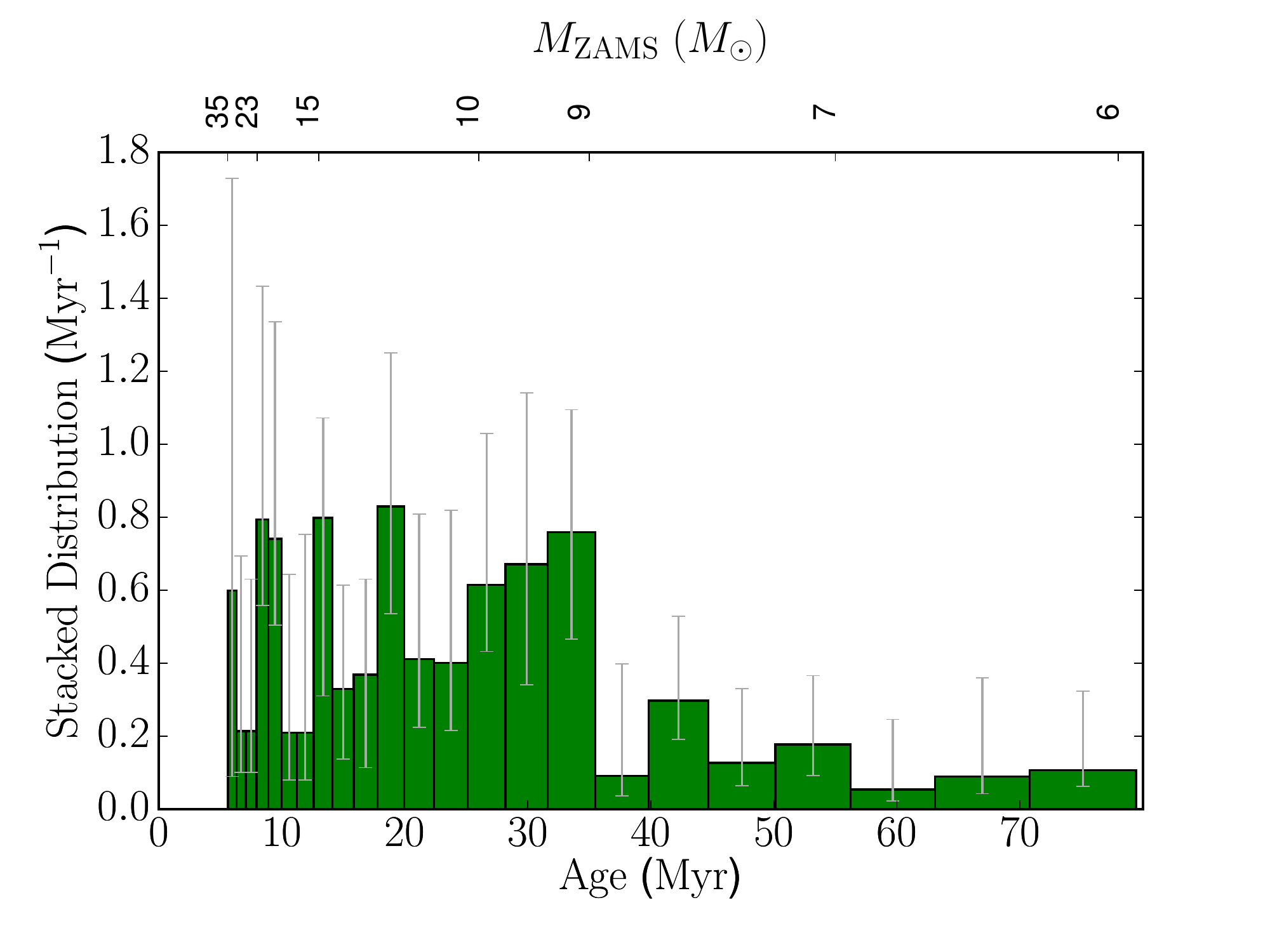}
	\caption{Stacked age distribution for the SN sample
		shown in Table~\ref{tab:historicsnetable}. For each historic SN,
		we derived a probability distribution from the individual
		SFHs; see Equation~\eqref{eq:probofburst}. We then sum those
		distributions to construct an age distribution for all SNe in our sample. Similar to \citet{diaz-rodriguez2018}, we
		model this distribution using two components.  Component one
		represents a power-law distribution with a slope of $\beta$, a
		minimum age, $\tmin$, and a maximum age, $\tmax$. In this
		analysis, we fix $\tmin$ due to the low number of SNe in this
		analysis and highly uncertain SFR for young age bins. The second component is an underlying uniform distribution that represents random unassociated bursts of star formation. }
	\label{fig:stackeddist}
\end{figure*}

Figure~\ref{fig:stackeddist}, shows the stacked age distribution for all $\nsn$ SNe. To produce this stacked distribution, we sum the PDFs for each SN ($\sum_k P_{k'}(i)$). In a previous analysis of 94 SNRs \citepalias{diaz-rodriguez2018}, the stacked age distribution clearly exhibited two distributions: one associated with the progenitors and one associated with random unassociated bursts of star formation. The stacked distribution of \citetalias{diaz-rodriguez2018} also showed similar features in the progenitor age distribution, namely a clear $\tmax$ (hence $\minmass$) and a clear smooth distribution
below that $\tmax$. While the number of SNe in this sample is lower than the \citetalias{diaz-rodriguez2018} study (e.g., $\nsn$ SNe vs. 94 SNRs), the stacked
distribution in Figure~\ref{fig:stackeddist} shows similar features to those in \citetalias{diaz-rodriguez2018}, although a little less clear. The most obvious feature is a reduction in the stacked distribution around 35-45 Myr; which likely reflects the maximum age (i.e., minimum mass) of the progenitor distribution. 

Section 2.4 of \citetalias{diaz-rodriguez2018} derived the progenitor age distribution model considering the SF in each age bin, $i$, as independent bursts of SF. For simplicity, \citetalias{diaz-rodriguez2018} considered a power-law distribution ($\frac{\mathrm{dN}}{\mathrm{dt}} \propto t^{\beta}$)
with a $\tmin$, $\tmax$, and slope ($\beta$) in between. Since the stacked distribution in Figure~\ref{fig:stackeddist} seems consistent with the stacked distribution in DR18, we adopt the same power-law model as in \citetalias{diaz-rodriguez2018}. The primary difference is that we set $\tmin$ to a fixed value in this study. Hence, our model only depends upon two parameters, $\tmax$ and power-law slope $\beta$. 

 We fixed the $\tmin$ parameter because the SFR for the younger age bins are highly uncertain, and these uncertainties are estimates based upon a {\hybridMC}
 \citep{dolphin2013}. The large uncertainties on the younger age bins limit the accuracy in inferring $\tmin$. The approximate nature of the {\hybridMC} compounds this problem. Hence, there is not enough information in this sample to effectively constrain both $\tmin$ and the power-law slope $\beta$ simultaneously. For simplicity, we only infer $\tmax$ and $\beta$ because we are more certain about the uncertainties for the higher age bins.

Figures~\ref{fig:w14historicsfhs} and \ref{fig:w18historicsfhs} often show multiple bursts of SF for each SN. Presumably one is associated with the progenitor and the other are random unassociated bursts of SF. Therefore, we use the likelihood model of \citetalias{diaz-rodriguez2018}, in which the true age of each SF burst, $\hat{t}$, is drawn from the power-law (progenitor) distribution or random uniform distribution, $P_p(\hat{t}|\theta)$ and $P_u(\hat{t})$ respectively. We defined the expressions for these two likelihoods in \citetalias{diaz-rodriguez2018} (see Equation 6 and 7 for more details). For each SN, the likelihood for the SFH is $\mathcal{L}_{k'}(\{ {\rm SFH} \}|\theta) = \mathcal{L}_{k'}(\{ i \}|\theta)$ and the final reduced likelihood is
\begin{equation}
	\label{eq:likelihoodfork3}
	\mathcal{L}_{k'}(\{ i\}|\theta) =
	C \cdot T_{\rm max} \cdot \sum_{j=1}^{N_\mathrm{SNe}} P_{k'}(j) \cdot P_p(j|\theta) \, ,
\end{equation}
see Section~2.5 of \citetalias{diaz-rodriguez2018} for a full description of this technique. Finally, the posterior distribution for the model parameters is proportional to: 
\begin{equation}
	P(\theta|\{ \rm{SFH} \}) \propto \mathcal{L}(\{ \rm{SFH} \}|\theta) \cdot P(\theta) \, .
\end{equation}



Using Equation~\ref{eq:posteriorjackknife} we infer the final posterior distribution, $P_r(\theta|\{\textrm{CMD}_{k^{\prime}}\})$ for each $r$th subsample using {\fontfamily{qcr}\selectfont emcee}. {\fontfamily{qcr}\selectfont Emcee} is a python implementation \citep{foreman-mackey2013,foreman-mackey2019} of the Affine Invariant Markov chain Monte Carlo Ensemble sampler by \citet{goodmanweare2010}. For this analysis, we use 100 walkers, 200 steps each, and we burn 100 of those. We use 200 steps because we noticed that it always converged to the most likely value using less than 100 steps. The resulting acceptance fraction for this inference is on average around $a_f=0.625$. The acceptance fraction represents the fraction of proposed steps that were accepted during the sampling \citep{foreman-mackey2013}.

Table~\ref{tab:priortable} shows the constraints for the prior distributions, $P(\theta)$.  For this study, the prior distribution is $P(\theta)~=~P(\tmax)~\cdot~P(\beta)$.

\begin{table}
	\begin{threeparttable}
		\centering
		\caption{Conditions for the prior distributions.}
		\label{tab:priortable}
		\begin{tabular}{lc}
			\toprule
			Parameter  & Prior\\
			\midrule
			Minimum age $\tmin$ & $\mathcal{U}( \historicminage \, {\rm
				Myr},\tmax)${\textbf{$^1$}}\\
			Maximum age $\tmax$ & $\mathcal{U}(\historicminage\, {\rm Myr},T_{\rm max})${\textbf{$^2$}} \\
			Slope $\beta$ &  $\mathcal{U}(-1,10)$ \\
			\bottomrule
		\end{tabular}
		\begin{tablenotes}
			\small
			\item {\textbf{$^1$}} The posterior distributions of $\tmax$ and $\beta$ are insensitive to the choice of $\tmin$. We fix $\tmin$ to $\historicminage$ Myr.
			\item {\textbf{$^2$}} We only considered the SNe with non-zero SF within the last $T_{\rm max} = 80$ Myr.
		\end{tablenotes}
	\end{threeparttable}
\end{table}

\section{Results}
\label{section:results}

\begin{figure}
	\centering
	\includegraphics[width=0.5\textwidth]{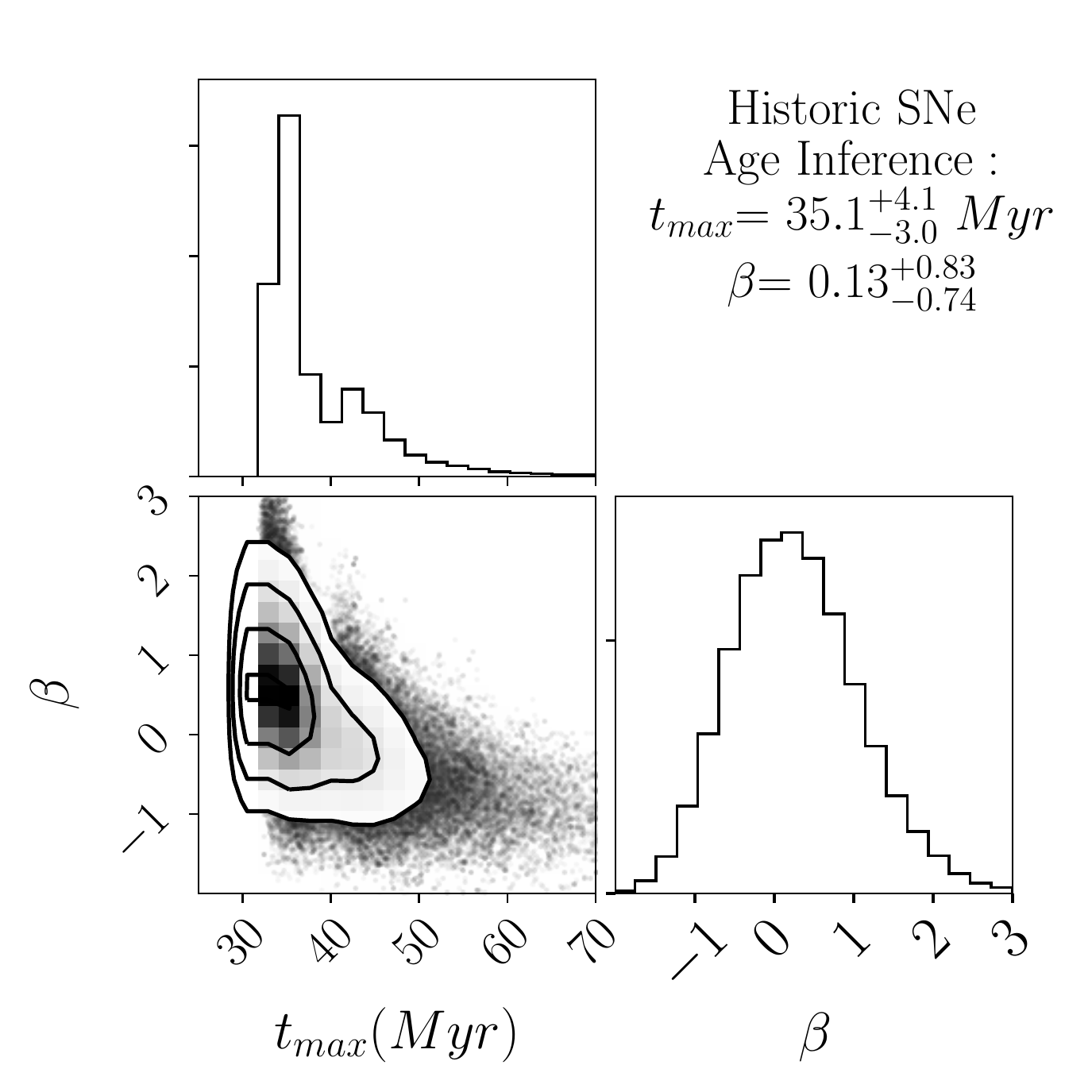}
	\caption{Posterior distribution for the model parameters of the
		age distribution: the maximum age is $\tmax$, and the power-law
		slope is $\beta$. We fixed the minimum age in our inference to
		be $\tmin$~=~$\historicminage$~Myr. The mode and narrowest 68\%
		confidence interval for the parameters are $\tmax= \historictmaxvalue$~Myr and $\beta=\historicbetavalue$.} 
	\label{fig:ageposterior}
\end{figure}

The posterior distribution for the maximum age, $\tmax$, and power-law slope for the age distribution, $\beta$, are shown in
Figure~\ref{fig:ageposterior}. While the proposed age distribution
model also depends upon a minimum age ($\tmin$), we do not infer its
value; rather, we fix it to $\historicminage$ Myr as discussed below.

\begin{figure*}
	\centering
	\subfloat{
		\includegraphics[width=1\columnwidth]{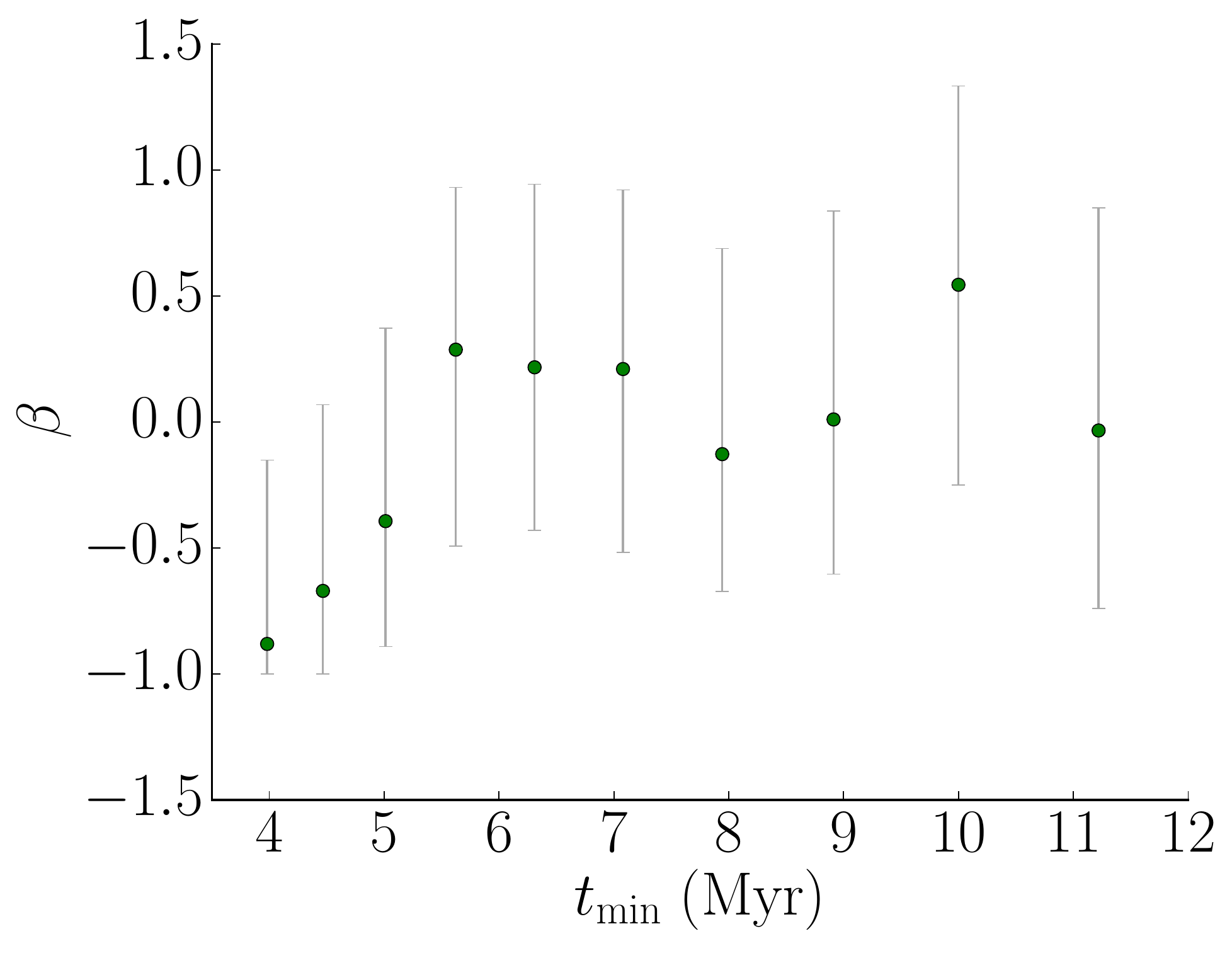}
		\label{fig:betavstmin}}
	\subfloat{
		\includegraphics[width=1\columnwidth]{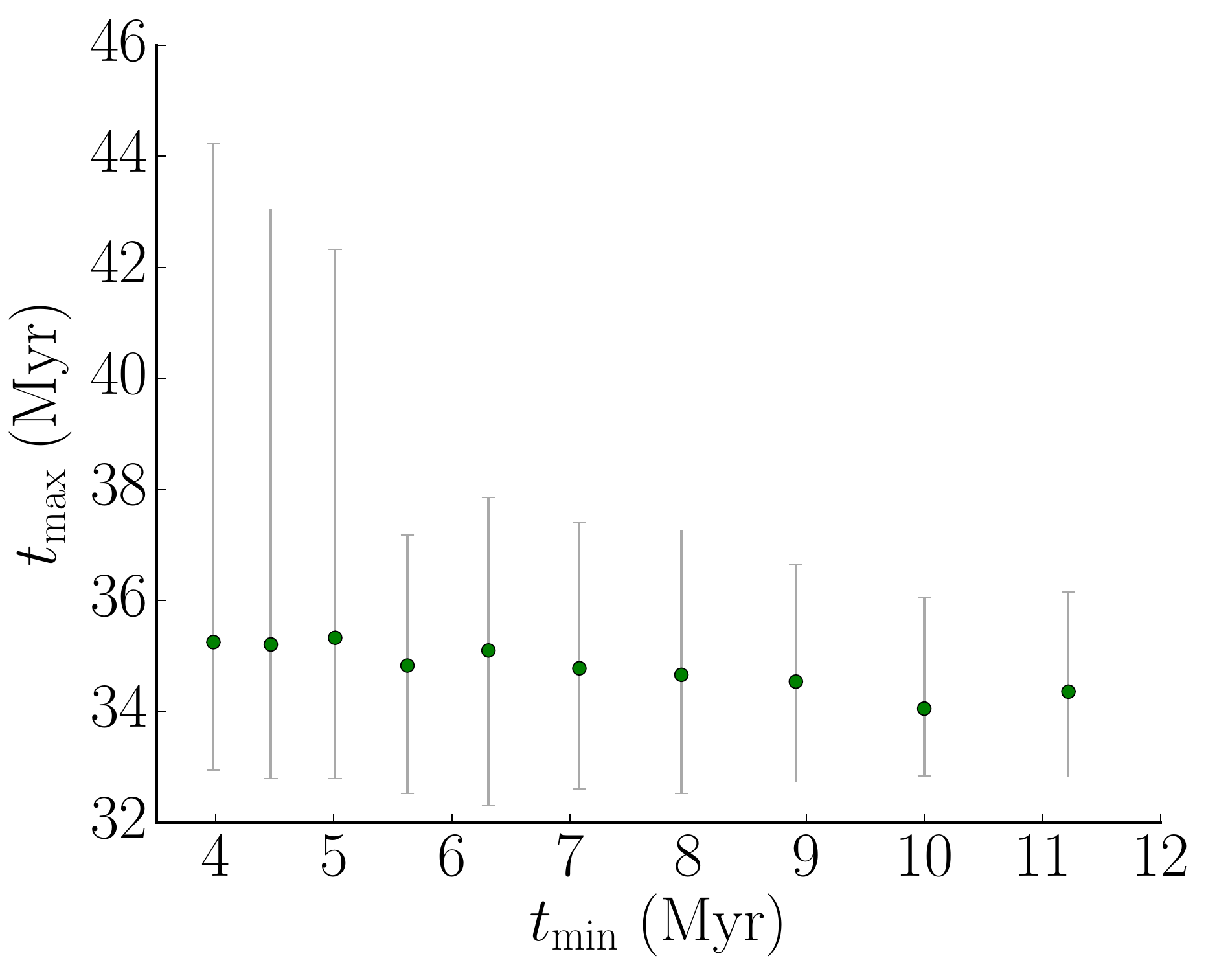}
		\label{fig:tmaxvstmin}}
	\caption{Inferred model parameters as a function of minimum age,
		$\tmin$. The green dots represent the mode for $\beta$ and $\tmax$, and the gray bars represent the narrowest 68\% confidence interval for the uncertainty. Below $\tmin~<~ \historicminage$~Myr the inference for $\beta$ is not stable; above it is.  Since the uncertainties are largest and least certain at the youngest ages, we restrict our analysis to age above $\historicminage$~Myr.  On the other
		hand, the inferences for $\tmax$ are mostly consistent for all trials of $\tmin$. For $\tmin <~ \historicminage$ Myr, the confidence interval for $\tmax$ are much larger. This is largely due to the uncertainties in the SFH at
		young age bins. For these reasons, we set $\tmin$ age bin to $\historicminage$ Myr for the rest of the analysis.}
	\label{fig:tmintest}
\end{figure*}

Figures~\ref{fig:betavstmin} and \ref{fig:tmaxvstmin} show $\beta$ and $\tmax$ inferences as a function of the minimum age, $\tmin$. The SFHs are calculated using logarithmic spaced age bins,
and the youngest edge of the minimum age bin correspond to
$\log_{10}(t/\text{yr}) = 6.6$. The right panel of Figure~\ref{fig:tmintest} shows that all inferences for $\tmax$ are consistent within the confidence intervals.  However, the confidence intervals for $\tmax$ are much larger for $\tmin < \historicminage$ Myr.  This is due to the large variation in SFH at the young age bins (see Figure~\ref{fig:w14historicsfhs} and
Figure~\ref{fig:w18historicsfhs}) and complexities of the evolution of massive stars. On the other hand, the inferences for $\beta$ (left panel in Figure~\ref{fig:tmintest}) show that $\beta$ is significantly different for inferences where $\tmin < \historicminage$ Myr.  Since we know that the older age bins are more certain than the younger age bins, this suggests that $\tmin$ should not be less than $\historicminage$ Myr. Hence, for the rest of the analysis, we fix
$\tmin$ to $\log_{10}(\tmin /\text{yr}) = \historiclogminage$; this is $\tmin =\historicminage$ Myr.

It is impossible to do a $\tmin$ sensitivity analysis with the SFH associated with SN192M and SN2005af.  In the range between $0-50$~Myr, both only have one burst of SF in the youngest age bin.  Omitting age bins younger than 5.62~Myr for those two cases and calculating the likelihood gives precisely zero. Therefore, it is impossible to do the $\tmin$ sensitivity study when including these two SNe. For that reason, we omit these two SNe for our entire analysis resulting in $N_\mathrm{SNe}=22$.

The marginalized values for the age distribution parameters are as follows. The $\tmax$ is $\historictmaxvalue$~Myr and the power-law slope for the age distribution is $\beta~=~\historicbetavalue$. The uncertainties reported for both parameters correspond to the narrowest 68 \% confidence interval.

\subsection{Age-to-$\mzams$}
\begin{figure}
	\centering
	\includegraphics[width=0.5\textwidth]{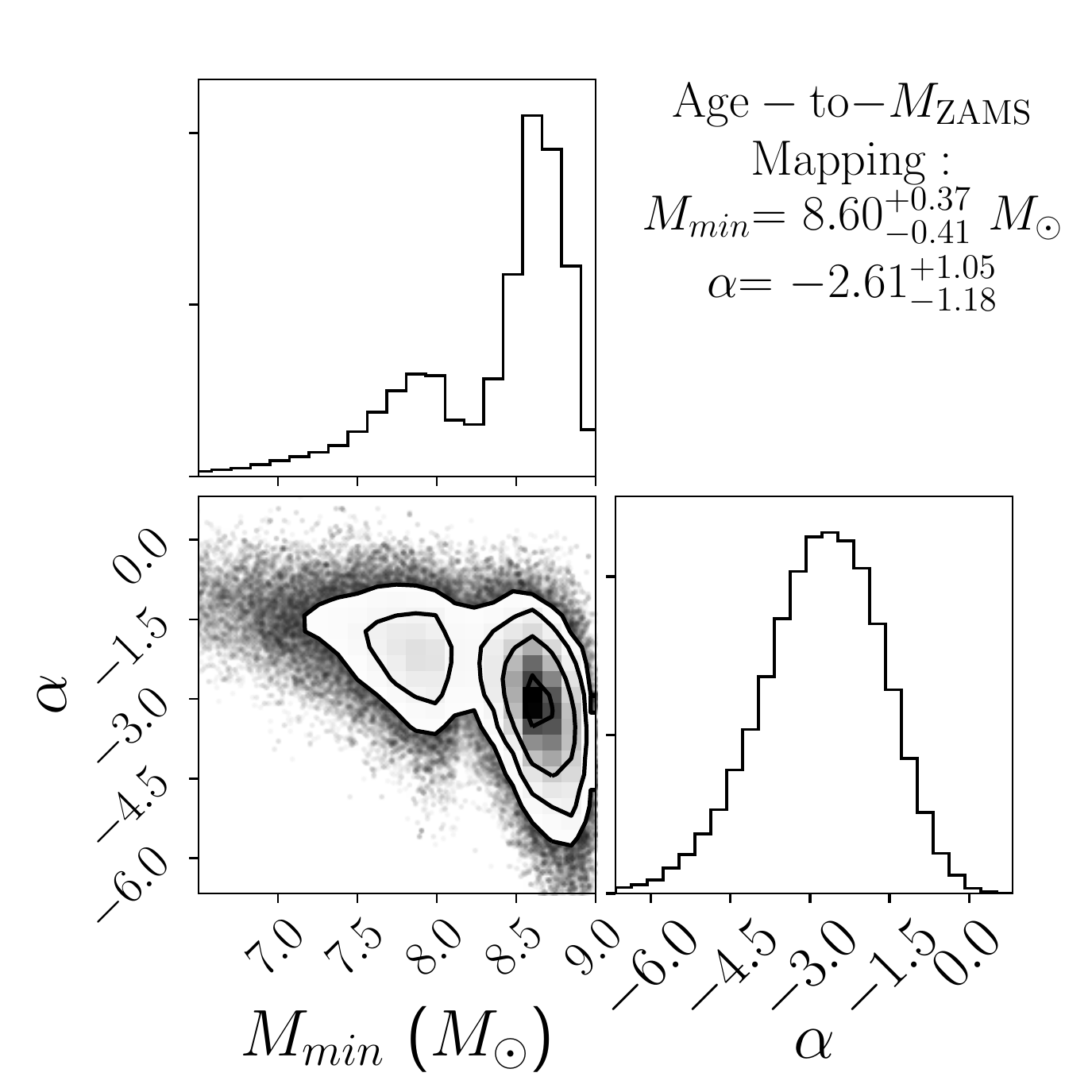}
	\caption{Posterior distribution for the progenitor-mass distribution parameters. Using a mapping between age and mass for single star evolution, we transform the posterior distribution for $\tmax$ and $\beta$ (Figure~\ref{fig:ageposterior}) to a posterior distribution for $\minmass$ and $\alpha$. The minimum mass is $\minmass=\historicminmassvalue \msun$ and the slope is $\alpha=\historicalphavalue$. The power-law index inferred in this analysis is consistent with the inference of \citet{diaz-rodriguez2018} which used 94 supernova remnants (SNRs), while the $\minmass$ is consistent within the 95\% confidence interval (see Figure~\ref{fig:massposteriorcomparison}). The large uncertainty for $\alpha$ is consistent with a wide range of distributions, including Salpeter and the steep power-law inferred from the SNRs.}
	\label{fig:massposterior}
\end{figure}

The main scientific goal for this paper is to understand which stars explode. However, our data is the SFH of the resolved stars within 50 pc around the location of each SN. Hence, we first infer the progenitor age distribution given the SFHs, which then must be translated into a progenitor mass distribution. We do this by applying an age-to-$\mzams$ mapping to obtain the progenitor mass distribution shown in Figure~\ref{fig:massposterior}. In mass space, $\tmax$ translates to a $\minmass$ and $\beta$ translates to a power-law slope of $\alpha$.

The mapping from age to mass requires a few assumptions. We assume solar metallicity and that these progenitors are single-stars. Hence, to map the progenitor age distribution to a progenitor mass distribution, we use
stellar evolution models from \citet{marigo2017}. To obtain the distributions for the $\minmass$ in Figure~\ref{fig:massposterior}, we first use the stellar evolution models to find an age vs. mass relationship; see Figure~8 of \citetalias{diaz-rodriguez2018}.  We then use this age vs. mass to find an $\minmass$ for each $\tmin$ in the MCMC step distribution.  To infer the power-law slope in mass, we use the chain rule to convert the age distribution to a mass distribution. In terms of $\beta$ (the power-law slope in age) and $\gamma$ (the power-law slope in the age to mass function), the power-law slope in mass is $\alpha=\historicalphavalue$.  To find the distribution of $\alpha$, we evaluate this formula for each $\beta$ in the MCMC chain. See section~3 of \citetalias{diaz-rodriguez2018} for details.

Assuming that all these historic SNe resulted from the explosion of
single stars, the marginalized mass distribution model parameters are $\minmass = \historicminmassvalue \ \msun$ and the slope of the progenitor distribution is $\alpha = \historicalphavalue$. The posterior distributions for the mass distribution parameters are in Figure~\ref{fig:massposterior}.

\section{Discussion \& Conclusion}
\label{section:discussionandconclusion}

Using a Bayesian hierarchical model previously derived in
\citetalias{diaz-rodriguez2018}, we infer the progenitor age distribution for
$\nsn$ historic CCSNe. We use the SFH for nearby stars ($r < 50 $pc) as a proxy for the age distribution for each SN. The SFHs
for 11 of the SNe were previously published in \citetalias{williams2014}, and the remaining 11 SFHs are from
\citetalias{williams2018}. We infer a $\tmax$ for CCSNe of
$\tmax=\historictmaxvalue$~Myr and a power-law slope for the progenitor age distribution of $\beta=\historicbetavalue$; while fixing the minimum age to $\tmin=\historicminage$~Myr (see Figure~\ref{fig:ageposterior}). Under the assumption that all progenitors were single stars, the age distribution translates to a progenitor mass distribution with a minimum mass of $\minmass=\historicminmassvalue \msun$ and a slope of $\alpha = \historicalphavalue$ (see Figure~\ref{fig:massposterior}). Both the $\minmass$ and power-law slope $\alpha$ are consistent within uncertainty with previous studies using the direct-imaging and age-dating techniques (see Figure~\ref{fig:massposteriorcomparison}). This study is also consistent with progenitor studies using metal abundance study of Galactic SNRs \citep{katsuda2018}.

\subsection{Minimum Mass for CCSNe}

Predicting the exact value for the $\minmass$ is not trivial and it depends upon a variety of factors such as
metallicity, binary interactions, mixing
processes, rotation, etc. (e.g., \citealt{eldridgetout2004, IbelingHeger2013, doherty2015, zapartas2017}). \citet{IbelingHeger2013} studied the minimum initial mass for classical CCSNe as a function of metallicity, $Z$, using non-rotating models. In this study, they found a mass limit of $\sim 8.35 \ \msun$ at $Z = -2.3$ and a limit of $\sim 9.5 \ \msun$ at $Z =0$. Moreover, they found that the initial mass limit continued to rise with higher metallicity. They also found that for a fixed initial mass function, the SN rate is 20\%~-~25\% higher at low metallicity. In addition to this, various studies have shown that the convective overshooting parameter can affect the minimum initial mass values for an SN. This extra mixing is used to simulate any mixing process and not just convective overshooting, like rotation or gravity wave mixing \citep{eldridgetout2004}. Essentially, a higher convective overshooting parameter and low metallicity will generally shift the $\minmass$ to lower values \citep{eldridgetout2004,Podsiadlowski2004}.

It is difficult to constrain the progenitors of CCSNe, because only a few of them have been discovered directly, making the statistical constraints quite loose.  \citetalias{williams2018} inferred the progenitor masses for 25 historic CCSNe (SN sample used in this work + 3 SN) using the SFHs near each SN. They found that the sample is consistent with a $\minmass$ for CCSNe of $<9.5 \ \msun$ (90\% confidence). In this study, we determine an initial mass of
$\minmass = \historicminmassvalue \ \msun$ for $\nsn$ historic
SNe. The RSG sample \citep{smartt2015} and the Historic SN sample from this study gives consistent results within a 68\% confidence interval (see Figure~\ref{fig:massposteriorcomparison}).  On the other hand, the SNR sample \citepalias{diaz-rodriguez2018} is only consistent at the 2-sigma level. More upcoming data will determine whether this slight inconsistency is a result due to a small sample size or some physical bias. If the inconsistency persists, then it could be due to several factors. For one, the $\minmass$ might be a function of metallicity as other studies have shown \citep[e.g.][]{ eldridgetout2004,Podsiadlowski2004}.  The Historic SN and RSG samples represent a heterogeneous metallicity sample, whereas, the SNR sample comes from two particular galaxies and presumably a narrower range of metallicity. Another possible explanation is that SNRs are biased toward older ages (hence smaller masses). Therefore, a definitive resolution will require more progenitor stars.
Binary evolution may also affect progenitor mass distribution inferences for
SNe \citep{demink2014merger}. There are two ways in which binary
evolution may affect age estimates.  For one, mass transfer from one
star to another can alter the relationship between zero-age-main
sequence mass and age. For instance, \citet{Zapartas2019} found that approximately 33\% to 50\% of Type II SN progenitors had exchange mass with a companion before exploding for most model assumptions. For example, two 4 $\msun$ stars could merge when one evolves off of the MS. The age of these stars and surrounding population would correspond to the age of a 4 $\msun$ star near its death, yet it would explode as an 8 $\msun$ star. In this scenario, the surrounding population is much older than the single star scenario would allow. In the second case, when one of the binary stars
explodes, the companion could be ejected at a high velocity. As a consequence, the star that was kicked out could eventually explode in an older region. 

If one does not account for binarity, the initial mass inferred for binary stars could be
overestimated by 15-25\% than its true initial mass \citep{Zapartas2020}. Further illustrating a need to consider binary evolution, the sample in this study contains four Type Ib/c SNe that have a wide range of ages \citepalias{williams2018}. Hence,
\citetalias{williams2018} suggested that this result is consistent with a large fraction of SN Ib/c originating from binary evolution. While these previous investigations
motivate a need for including binary evolution in inferring ages, we use single star models in this analysis to provide a baseline for comparison.  

\subsection{Mass Distribution Power-Law Slope}

\begin{figure}
	\centering
	\includegraphics[width=0.5\textwidth]{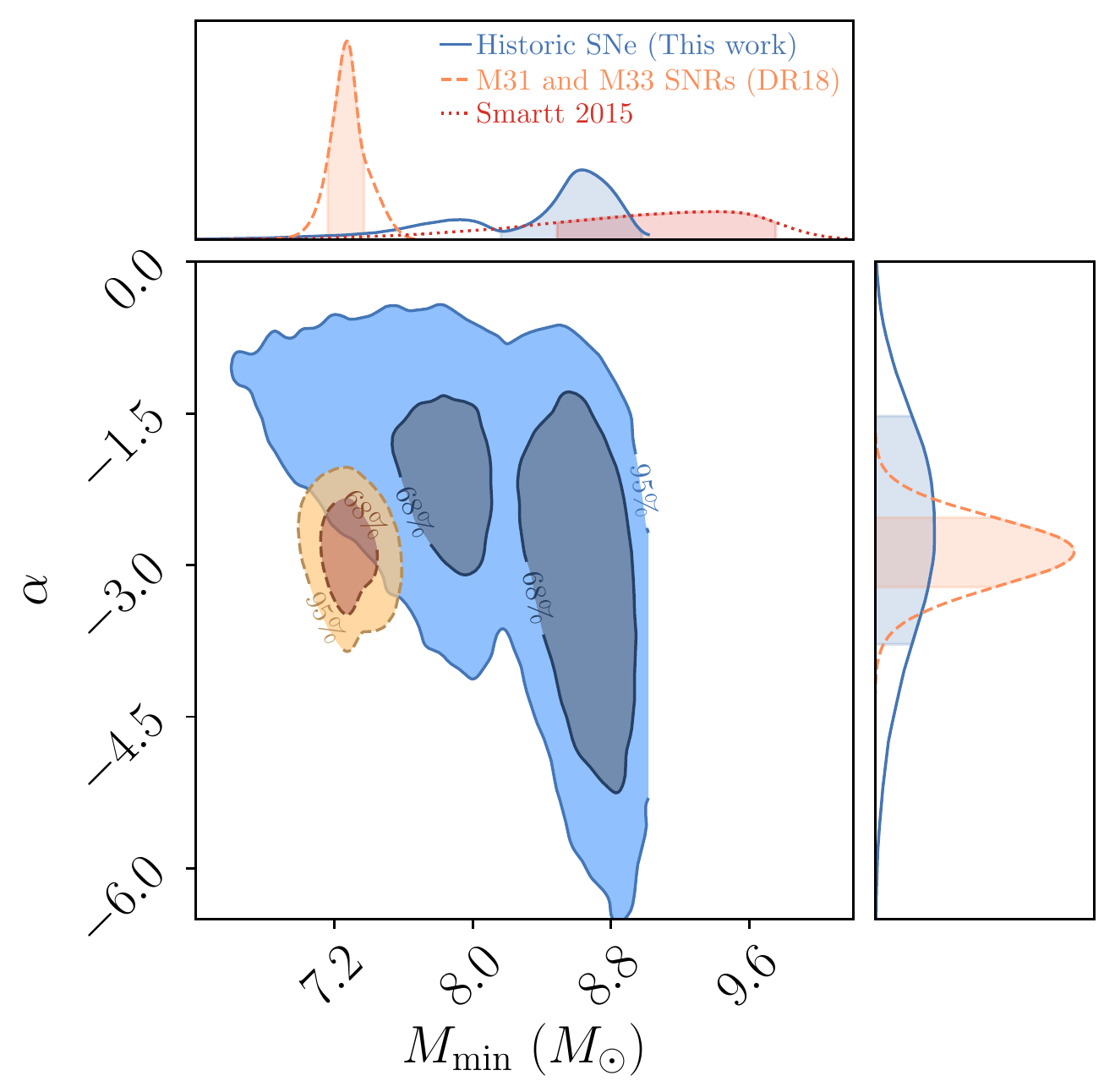}
	\caption{Posterior distribution for the progenitor-mass distribution parameters. This figure compares the results for this work (Figure~\ref{fig:massposterior}, blue contour) with previous progenitor mass distribution inferences. The orange contour shows the results of \citetalias{diaz-rodriguez2018}, which inferred the distribution using 94 supernova remnants for M31 and M33. The red distribution in the top panel shows the distribution from 26 type II SN (mostly type II-P) \citep{smartt2015}. The power-law index ($\alpha$) for both this and the \citetalias{diaz-rodriguez2018} studies are consistent with each other, while the $\minmass$ inferred for the SNRs is slightly lower than the inference using the historic SNe. The \citet{smartt2015} analysis has relatively large uncertainties and is formally consistent with both age-dating inferences.}
	\label{fig:massposteriorcomparison}
\end{figure}

We find a power-law slope of $\alpha= \historicalphavalue$ for the
progenitor mass distribution. This value is consistent with the standard Salpeter IMF ($dN /
dM \propto M^{\alpha} \mathrm{, where}$ $\alpha = -2.35$). Both our
inferred parameters, $\minmass$ and $\alpha$, are consistent within
the uncertainties of previous progenitor
estimates using direct detections and the age-dating technique
\citep{smartt2015,jennings2014,williams2018}. \citetalias{williams2018}
estimated the masses for 25 CCSNe ($\nsn$ of these are consider in this
study) and found that the progenitor mass distribution is consistent
with a Salpeter IMF distribution as well. \citetalias{diaz-rodriguez2018} found a steeper
slope of $\alpha = -2.96^{+0.45}_{-0.25}$ using 94 SNRs in M31 and
M33 (see Figure~\ref{fig:massposteriorcomparison}). The results of \citetalias{diaz-rodriguez2018} suggest that either the most massive stars are not exploding as frequently as lower mass stars, or there is a bias against observing SNRs in the youngest regions. While those results prefer a slight steeper
distribution, and the results of this paper are consistent with a
Salpeter IMF, both are consistent with one another.  For the most
part, this is because the confidence interval of this study is larger.

\section*{Acknowledgements}
Based on observations made with the NASA/ESA Hubble Space Telescope, obtained [from the Data Archive] at the Space Telescope Science Institute, which is operated by the Association of Universities for Research in Astronomy, Inc., under NASA contract NAS 5-26555. Support for programs \# HST-AR-13882, \# HST-AR-15042, and \# HST-GO-14786 was provided by NASA through a grant from the Space Telescope Science Institute, which is operated by the Association of Universities for Research in Astronomy, Inc., under NASA contract NAS 5-26555.




\bibliographystyle{mnras}
\bibliography{myrefs}

\begin{thebibliography}{}
\makeatletter
\relax
\def\mn@urlcharsother{\let\do\@makeother \do\$\do\&\do\#\do\^\do\_\do\%\do\~}
\def\mn@doi{\begingroup\mn@urlcharsother \@ifnextchar [ {\mn@doi@}
  {\mn@doi@[]}}
\def\mn@doi@[#1]#2{\def\@tempa{#1}\ifx\@tempa\@empty \href
  {http://dx.doi.org/#2} {doi:#2}\else \href {http://dx.doi.org/#2} {#1}\fi
  \endgroup}
\def\mn@eprint#1#2{\mn@eprint@#1:#2::\@nil}
\def\mn@eprint@arXiv#1{\href {http://arxiv.org/abs/#1} {{\tt arXiv:#1}}}
\def\mn@eprint@dblp#1{\href {http://dblp.uni-trier.de/rec/bibtex/#1.xml}
  {dblp:#1}}
\def\mn@eprint@#1:#2:#3:#4\@nil{\def\@tempa {#1}\def\@tempb {#2}\def\@tempc
  {#3}\ifx \@tempc \@empty \let \@tempc \@tempb \let \@tempb \@tempa \fi \ifx
  \@tempb \@empty \def\@tempb {arXiv}\fi \@ifundefined
  {mn@eprint@\@tempb}{\@tempb:\@tempc}{\expandafter \expandafter \csname
  mn@eprint@\@tempb\endcsname \expandafter{\@tempc}}}

\bibitem[\protect\citeauthoryear{{Auchettl}, {Lopez}, {Badenes},
  {Ramirez-Ruiz}, {Beacom}  \& {Holland -Ashford}}{{Auchettl}
  et~al.}{2019}]{Auchettl2019}
{Auchettl} K.,  {Lopez} L.~A.,  {Badenes} C.,  {Ramirez-Ruiz} E.,  {Beacom}
  J.~F.,   {Holland -Ashford} T.,  2019, \mn@doi [\apj]
  {10.3847/1538-4357/aaf395}, \href
  {https://ui.adsabs.harvard.edu/abs/2019ApJ...871...64A} {871, 64}

\bibitem[\protect\citeauthoryear{Badenes, Harris, Zaritsky  \& Prieto}{Badenes
  et~al.}{2009}]{badenes2009}
Badenes C.,  Harris J.,  Zaritsky D.,   Prieto J.~L.,  2009, \mn@doi [The
  Astrophysical Journal] {10.1088/0004-637x/700/1/727}, 700, 727

\bibitem[\protect\citeauthoryear{{Badenes}, {Maoz}  \& {Ciardullo}}{{Badenes}
  et~al.}{2015}]{badenes2015}
{Badenes} C.,  {Maoz} D.,   {Ciardullo} R.,  2015, \mn@doi [\apjl]
  {10.1088/2041-8205/804/1/L25}, \href
  {https://ui.adsabs.harvard.edu/abs/2015ApJ...804L..25B} {804, L25}

\bibitem[\protect\citeauthoryear{{Beasor} \& {Davies}}{{Beasor} \&
  {Davies}}{2016}]{beasor2016}
{Beasor} E.~R.,  {Davies} B.,  2016, \mn@doi [\mnras] {10.1093/mnras/stw2054},
  \href {http://adsabs.harvard.edu/abs/2016MNRAS.463.1269B} {463, 1269}

\bibitem[\protect\citeauthoryear{{Bovy}}{{Bovy}}{2016}]{Bovy2016}
{Bovy} J.,  2016, \mn@doi [\apj] {10.3847/0004-637X/817/1/49}, \href
  {https://ui.adsabs.harvard.edu/abs/2016ApJ...817...49B} {817, 49}

\bibitem[\protect\citeauthoryear{{Burrows} \& {Goshy}}{{Burrows} \&
  {Goshy}}{1993}]{burrows93}
{Burrows} A.,  {Goshy} J.,  1993, \mn@doi [\apjl] {10.1086/187074}, \href
  {http://adsabs.harvard.edu/abs/1993ApJ...416L..75B} {416, L75}

\bibitem[\protect\citeauthoryear{Colquhoun}{Colquhoun}{2014}]{colquhoun2014}
Colquhoun D.,  2014, Royal Society open science, 1, 140216

\bibitem[\protect\citeauthoryear{{Crockett} et~al.,}{{Crockett}
  et~al.}{2008}]{crockett2008}
{Crockett} R.~M.,  et~al., 2008, \mn@doi [\mnras]
  {10.1111/j.1745-3933.2008.00540.x}, \href
  {http://adsabs.harvard.edu/abs/2008MNRAS.391L...5C} {391, L5}

\bibitem[\protect\citeauthoryear{Davies \& Beasor}{Davies \&
  Beasor}{2018}]{daviesbeasor2018}
Davies B.,  Beasor E.~R.,  2018, \mn@doi [Monthly Notices of the Royal
  Astronomical Society] {10.1093/mnras/stx2734}, 474, 2116

\bibitem[\protect\citeauthoryear{{Davies} \& {Beasor}}{{Davies} \&
  {Beasor}}{2020}]{DaviesandBeasor2020}
{Davies} B.,  {Beasor} E.~R.,  2020, \mn@doi [\mnras] {10.1093/mnras/staa174},
  \href {https://ui.adsabs.harvard.edu/abs/2020MNRAS.493..468D} {493, 468}

\bibitem[\protect\citeauthoryear{{D{\'{\i}}az-Rodr{\'{\i}}guez}, {Murphy},
  {Rubin}, {Dolphin}, {Williams}  \&
  {Dalcanton}}{{D{\'{\i}}az-Rodr{\'{\i}}guez}
  et~al.}{2018}]{diaz-rodriguez2018}
{D{\'{\i}}az-Rodr{\'{\i}}guez} M.,  {Murphy} J.~W.,  {Rubin} D.~A.,  {Dolphin}
  A.~E.,  {Williams} B.~F.,   {Dalcanton} J.~J.,  2018, \mn@doi [\apj]
  {10.3847/1538-4357/aac6e1}, \href
  {http://adsabs.harvard.edu/abs/2018ApJ...861...92D} {861, 92}

\bibitem[\protect\citeauthoryear{{Doherty}, {Gil-Pons}, {Siess}, {Lattanzio}
  \& {Lau}}{{Doherty} et~al.}{2015}]{doherty2015}
{Doherty} C.~L.,  {Gil-Pons} P.,  {Siess} L.,  {Lattanzio} J.~C.,   {Lau} H.
  H.~B.,  2015, \mn@doi [\mnras] {10.1093/mnras/stu2180}, \href
  {https://ui.adsabs.harvard.edu/abs/2015MNRAS.446.2599D} {446, 2599}

\bibitem[\protect\citeauthoryear{{Dolphin}}{{Dolphin}}{2000}]{dolphin2000}
{Dolphin} A.~E.,  2000, \mn@doi [\pasp] {10.1086/316630}, \href
  {https://ui.adsabs.harvard.edu/abs/2000PASP..112.1383D} {112, 1383}

\bibitem[\protect\citeauthoryear{{Dolphin}}{{Dolphin}}{2002}]{dolphin2002}
{Dolphin} A.~E.,  2002, \mn@doi [\mnras] {10.1046/j.1365-8711.2002.05271.x},
  \href {http://adsabs.harvard.edu/abs/2002MNRAS.332...91D} {332, 91}

\bibitem[\protect\citeauthoryear{{Dolphin}}{{Dolphin}}{2012}]{dolphin2012}
{Dolphin} A.~E.,  2012, \mn@doi [\apj] {10.1088/0004-637X/751/1/60}, \href
  {http://adsabs.harvard.edu/abs/2012ApJ...751...60D} {751, 60}

\bibitem[\protect\citeauthoryear{{Dolphin}}{{Dolphin}}{2013}]{dolphin2013}
{Dolphin} A.~E.,  2013, \mn@doi [\apj] {10.1088/0004-637X/775/1/76}, \href
  {http://adsabs.harvard.edu/abs/2013ApJ...775...76D} {775, 76}

\bibitem[\protect\citeauthoryear{{Dolphin}}{{Dolphin}}{2016}]{dolphin2016}
{Dolphin} A.,  2016, {DOLPHOT: Stellar photometry}, Astrophysics Source Code
  Library (\mn@eprint {ascl} {1608.013})

\bibitem[\protect\citeauthoryear{{Dolphin} et~al.,}{{Dolphin}
  et~al.}{2003}]{dolphin2003}
{Dolphin} A.~E.,  et~al., 2003, \mn@doi [\aj] {10.1086/375761}, \href
  {https://ui.adsabs.harvard.edu/abs/2003AJ....126..187D} {126, 187}

\bibitem[\protect\citeauthoryear{Ebinger, Curtis, Fröhlich, Hempel, Perego,
  Liebendörfer  \& Thielemann}{Ebinger et~al.}{2018}]{ebinger2018}
Ebinger K.,  Curtis S.,  Fröhlich C.,  Hempel M.,  Perego A.,  Liebendörfer
  M.,   Thielemann F.-K.,  2018, \mn@doi [The Astrophysical Journal]
  {10.3847/1538-4357/aae7c9}, 870, 1

\bibitem[\protect\citeauthoryear{{Eldridge} \& {Tout}}{{Eldridge} \&
  {Tout}}{2004}]{eldridgetout2004}
{Eldridge} J.~J.,  {Tout} C.~A.,  2004, \mn@doi [\mnras]
  {10.1111/j.1365-2966.2004.08041.x}, \href
  {https://ui.adsabs.harvard.edu/abs/2004MNRAS.353...87E} {353, 87}

\bibitem[\protect\citeauthoryear{{Farrell}, {Groh}, {Meynet}  \&
  {Eldridge}}{{Farrell} et~al.}{2020}]{farrell2020}
{Farrell} E.~J.,  {Groh} J.~H.,  {Meynet} G.,   {Eldridge} J.~J.,  2020,
  \mn@doi [\mnras] {10.1093/mnrasl/slaa035}, \href
  {https://ui.adsabs.harvard.edu/abs/2020MNRAS.494L..53F} {494, L53}

\bibitem[\protect\citeauthoryear{{Foreman-Mackey}, {Hogg}, {Lang}  \&
  {Goodman}}{{Foreman-Mackey} et~al.}{2013}]{foreman-mackey2013}
{Foreman-Mackey} D.,  {Hogg} D.~W.,  {Lang} D.,   {Goodman} J.,  2013, \mn@doi
  [\pasp] {10.1086/670067}, \href
  {http://adsabs.harvard.edu/abs/2013PASP..125..306F} {125, 306}

\bibitem[\protect\citeauthoryear{{Foreman-Mackey} et~al.,}{{Foreman-Mackey}
  et~al.}{2019}]{foreman-mackey2019}
{Foreman-Mackey} D.,  et~al., 2019, \mn@doi [The Journal of Open Source
  Software] {10.21105/joss.01864}, \href
  {https://ui.adsabs.harvard.edu/abs/2019JOSS....4.1864F} {4, 1864}

\bibitem[\protect\citeauthoryear{{Fraser} et~al.,}{{Fraser}
  et~al.}{2014}]{fraser2014}
{Fraser} M.,  et~al., 2014, \mn@doi [\mnras] {10.1093/mnrasl/slt179}, \href
  {http://adsabs.harvard.edu/abs/2014MNRAS.439L..56F} {439, L56}

\bibitem[\protect\citeauthoryear{{Gal-Yam} \& {Leonard}}{{Gal-Yam} \&
  {Leonard}}{2009}]{galyam2009}
{Gal-Yam} A.,  {Leonard} D.~C.,  2009, \mn@doi [\nat] {10.1038/nature07934},
  \href {http://adsabs.harvard.edu/abs/2009Natur.458..865G} {458, 865}

\bibitem[\protect\citeauthoryear{{Gallart}, {Zoccali}  \& {Aparicio}}{{Gallart}
  et~al.}{2005}]{gallart2005}
{Gallart} C.,  {Zoccali} M.,   {Aparicio} A.,  2005, \mn@doi [\araa]
  {10.1146/annurev.astro.43.072103.150608}, \href
  {https://ui.adsabs.harvard.edu/abs/2005ARA&A..43..387G} {43, 387}

\bibitem[\protect\citeauthoryear{{Girardi} et~al.,}{{Girardi}
  et~al.}{2010}]{girardi2010}
{Girardi} L.,  et~al., 2010, \mn@doi [\apj] {10.1088/0004-637X/724/2/1030},
  \href {http://adsabs.harvard.edu/abs/2010ApJ...724.1030G} {724, 1030}

\bibitem[\protect\citeauthoryear{{Gogarten}, {Dalcanton}, {Murphy}, {Williams},
  {Gilbert}  \& {Dolphin}}{{Gogarten} et~al.}{2009}]{gogarten2009}
{Gogarten} S.~M.,  {Dalcanton} J.~J.,  {Murphy} J.~W.,  {Williams} B.~F.,
  {Gilbert} K.,   {Dolphin} A.,  2009, \mn@doi [\apj]
  {10.1088/0004-637X/703/1/300}, \href
  {http://adsabs.harvard.edu/abs/2009ApJ...703..300G} {703, 300}

\bibitem[\protect\citeauthoryear{{Goodman} \& {Weare}}{{Goodman} \&
  {Weare}}{2010}]{goodmanweare2010}
{Goodman} J.,  {Weare} J.,  2010, \mn@doi [Communications in Applied
  Mathematics and Computational Science, Vol.~5, No.~1, p.~65-80, 2010]
  {10.2140/camcos.2010.5.65}, \href
  {http://adsabs.harvard.edu/abs/2010CAMCS...5...65G} {5, 65}

\bibitem[\protect\citeauthoryear{{Hendry} et~al.,}{{Hendry}
  et~al.}{2006}]{hendry2006}
{Hendry} M.~A.,  et~al., 2006, \mn@doi [\mnras]
  {10.1111/j.1365-2966.2006.10374.x}, \href
  {http://adsabs.harvard.edu/abs/2006MNRAS.369.1303H} {369, 1303}

\bibitem[\protect\citeauthoryear{{Ibeling} \& {Heger}}{{Ibeling} \&
  {Heger}}{2013}]{IbelingHeger2013}
{Ibeling} D.,  {Heger} A.,  2013, \mn@doi [\apjl]
  {10.1088/2041-8205/765/2/L43}, \href
  {https://ui.adsabs.harvard.edu/abs/2013ApJ...765L..43I} {765, L43}

\bibitem[\protect\citeauthoryear{{Jennings}, {Williams}, {Murphy}, {Dalcanton},
  {Gilbert}, {Dolphin}, {Fouesneau}  \& {Weisz}}{{Jennings}
  et~al.}{2012}]{jennings2012}
{Jennings} Z.~G.,  {Williams} B.~F.,  {Murphy} J.~W.,  {Dalcanton} J.~J.,
  {Gilbert} K.~M.,  {Dolphin} A.~E.,  {Fouesneau} M.,   {Weisz} D.~R.,  2012,
  \mn@doi [\apj] {10.1088/0004-637X/761/1/26}, \href
  {http://adsabs.harvard.edu/abs/2012ApJ...761...26J} {761, 26}

\bibitem[\protect\citeauthoryear{{Jennings}, {Williams}, {Murphy}, {Dalcanton},
  {Gilbert}, {Dolphin}, {Weisz}  \& {Fouesneau}}{{Jennings}
  et~al.}{2014}]{jennings2014}
{Jennings} Z.~G.,  {Williams} B.~F.,  {Murphy} J.~W.,  {Dalcanton} J.~J.,
  {Gilbert} K.~M.,  {Dolphin} A.~E.,  {Weisz} D.~R.,   {Fouesneau} M.,  2014,
  \mn@doi [\apj] {10.1088/0004-637X/795/2/170}, \href
  {http://adsabs.harvard.edu/abs/2014ApJ...795..170J} {795, 170}

\bibitem[\protect\citeauthoryear{{Katsuda}, {Takiwaki}, {Tominaga}, {Moriya}
  \& {Nakamura}}{{Katsuda} et~al.}{2018}]{katsuda2018}
{Katsuda} S.,  {Takiwaki} T.,  {Tominaga} N.,  {Moriya} T.~J.,   {Nakamura} K.,
   2018, \mn@doi [\apj] {10.3847/1538-4357/aad2d8}, \href
  {https://ui.adsabs.harvard.edu/abs/2018ApJ...863..127K} {863, 127}

\bibitem[\protect\citeauthoryear{{Kochanek}}{{Kochanek}}{2020}]{Kochanek2020}
{Kochanek} C.~S.,  2020, \mn@doi [\mnras] {10.1093/mnras/staa605}, \href
  {https://ui.adsabs.harvard.edu/abs/2020MNRAS.493.4945K} {493, 4945}

\bibitem[\protect\citeauthoryear{{Lada} \& {Lada}}{{Lada} \&
  {Lada}}{2003}]{lada2003}
{Lada} C.~J.,  {Lada} E.~A.,  2003, \mn@doi [\araa]
  {10.1146/annurev.astro.41.011802.094844}, \href
  {http://adsabs.harvard.edu/abs/2003ARA%26A..41...57L} {41, 57}

\bibitem[\protect\citeauthoryear{{Leonard}}{{Leonard}}{2010}]{leonard2010}
{Leonard} D.~C.,  2010, in {Leitherer} C.,  {Bennett} P.~D.,  {Morris} P.~W.,
  {Van Loon} J.~T.,  eds,  Astronomical Society of the Pacific Conference
  Series Vol. 425, Hot and Cool: Bridging Gaps in Massive Star Evolution. p.~79
  (\mn@eprint {arXiv} {0908.1812})

\bibitem[\protect\citeauthoryear{{Lewis} et~al.,}{{Lewis}
  et~al.}{2015}]{lewis2015}
{Lewis} A.~R.,  et~al., 2015, \mn@doi [\apj] {10.1088/0004-637X/805/2/183},
  \href {http://adsabs.harvard.edu/abs/2015ApJ...805..183L} {805, 183}

\bibitem[\protect\citeauthoryear{{Li}, {Van Dyk}, {Filippenko}  \&
  {Cuillandre}}{{Li} et~al.}{2005}]{li2005}
{Li} W.,  {Van Dyk} S.~D.,  {Filippenko} A.~V.,   {Cuillandre} J.-C.,  2005,
  \mn@doi [\pasp] {10.1086/428278}, \href
  {http://adsabs.harvard.edu/abs/2005PASP..117..121L} {117, 121}

\bibitem[\protect\citeauthoryear{{Li}, {Van Dyk}, {Filippenko}, {Cuillandre},
  {Jha}, {Bloom}, {Riess}  \& {Livio}}{{Li} et~al.}{2006}]{li2006}
{Li} W.,  {Van Dyk} S.~D.,  {Filippenko} A.~V.,  {Cuillandre} J.-C.,  {Jha} S.,
   {Bloom} J.~S.,  {Riess} A.~G.,   {Livio} M.,  2006, \mn@doi [\apj]
  {10.1086/499916}, \href {http://adsabs.harvard.edu/abs/2006ApJ...641.1060L}
  {641, 1060}

\bibitem[\protect\citeauthoryear{{Li}, {Wang}, {Van Dyk}, {Cuillandre}, {Foley}
   \& {Filippenko}}{{Li} et~al.}{2007}]{li2007}
{Li} W.,  {Wang} X.,  {Van Dyk} S.~D.,  {Cuillandre} J.-C.,  {Foley} R.~J.,
  {Filippenko} A.~V.,  2007, \mn@doi [\apj] {10.1086/516747}, \href
  {http://adsabs.harvard.edu/abs/2007ApJ...661.1013L} {661, 1013}

\bibitem[\protect\citeauthoryear{{Mabanta} \& {Murphy}}{{Mabanta} \&
  {Murphy}}{2018}]{mabanta2018}
{Mabanta} Q.~A.,  {Murphy} J.~W.,  2018, \mn@doi [\apj]
  {10.3847/1538-4357/aaaec7}, \href
  {http://adsabs.harvard.edu/abs/2018ApJ...856...22M} {856, 22}

\bibitem[\protect\citeauthoryear{{Ma{\'{\i}}z-Apell{\'a}niz}, {Bond}, {Siegel},
  {Lipkin}, {Maoz}, {Ofek}  \& {Poznanski}}{{Ma{\'{\i}}z-Apell{\'a}niz}
  et~al.}{2004}]{maizapellaniz2004}
{Ma{\'{\i}}z-Apell{\'a}niz} J.,  {Bond} H.~E.,  {Siegel} M.~H.,  {Lipkin} Y.,
  {Maoz} D.,  {Ofek} E.~O.,   {Poznanski} D.,  2004, \mn@doi [\apjl]
  {10.1086/426120}, \href {http://adsabs.harvard.edu/abs/2004ApJ...615L.113M}
  {615, L113}

\bibitem[\protect\citeauthoryear{{Marigo} et~al.,}{{Marigo}
  et~al.}{2017}]{marigo2017}
{Marigo} P.,  et~al., 2017, \mn@doi [\apj] {10.3847/1538-4357/835/1/77}, \href
  {http://adsabs.harvard.edu/abs/2017ApJ...835...77M} {835, 77}

\bibitem[\protect\citeauthoryear{{Maund}}{{Maund}}{2017}]{maund2017}
{Maund} J.~R.,  2017, \mn@doi [\mnras] {10.1093/mnras/stx879}, \href
  {http://adsabs.harvard.edu/abs/2017MNRAS.469.2202M} {469, 2202}

\bibitem[\protect\citeauthoryear{{Maund}}{{Maund}}{2018}]{maund2018}
{Maund} J.~R.,  2018, \mn@doi [\mnras] {10.1093/mnras/sty093}, \href
  {http://adsabs.harvard.edu/abs/2018MNRAS.476.2629M} {476, 2629}

\bibitem[\protect\citeauthoryear{{Maund}, {Smartt}, {Kudritzki},
  {Podsiadlowski}  \& {Gilmore}}{{Maund} et~al.}{2004}]{maund2004}
{Maund} J.~R.,  {Smartt} S.~J.,  {Kudritzki} R.~P.,  {Podsiadlowski} P.,
  {Gilmore} G.~F.,  2004, \mn@doi [\nat] {10.1038/nature02161}, \href
  {https://ui.adsabs.harvard.edu/abs/2004Natur.427..129M} {427, 129}

\bibitem[\protect\citeauthoryear{{Maund}, {Smartt}  \& {Danziger}}{{Maund}
  et~al.}{2005}]{maundsn2005cs}
{Maund} J.~R.,  {Smartt} S.~J.,   {Danziger} I.~J.,  2005, \mn@doi [\mnras]
  {10.1111/j.1745-3933.2005.00100.x}, \href
  {http://adsabs.harvard.edu/abs/2005MNRAS.364L..33M} {364, L33}

\bibitem[\protect\citeauthoryear{{Maund} et~al.,}{{Maund}
  et~al.}{2011}]{maund2011}
{Maund} J.~R.,  et~al., 2011, \mn@doi [\apjl] {10.1088/2041-8205/739/2/L37},
  \href {http://adsabs.harvard.edu/abs/2011ApJ...739L..37M} {739, L37}

\bibitem[\protect\citeauthoryear{{Maund}, {Mattila}, {Ramirez-Ruiz}  \&
  {Eldridge}}{{Maund} et~al.}{2014}]{maund2014bk}
{Maund} J.~R.,  {Mattila} S.,  {Ramirez-Ruiz} E.,   {Eldridge} J.~J.,  2014,
  \mn@doi [\mnras] {10.1093/mnras/stt2296}, \href
  {http://adsabs.harvard.edu/abs/2014MNRAS.438.1577M} {438, 1577}

\bibitem[\protect\citeauthoryear{{Murphy} \& {Burrows}}{{Murphy} \&
  {Burrows}}{2008}]{murphy2008}
{Murphy} J.~W.,  {Burrows} A.,  2008, \mn@doi [\apj] {10.1086/592214}, \href
  {http://adsabs.harvard.edu/abs/2008ApJ...688.1159M} {688, 1159}

\bibitem[\protect\citeauthoryear{{Murphy} \& {Dolence}}{{Murphy} \&
  {Dolence}}{2017}]{murphy2017}
{Murphy} J.~W.,  {Dolence} J.~C.,  2017, \mn@doi [\apj]
  {10.3847/1538-4357/834/2/183}, \href
  {http://adsabs.harvard.edu/abs/2017ApJ...834..183M} {834, 183}

\bibitem[\protect\citeauthoryear{{Murphy}, {Jennings}, {Williams}, {Dalcanton}
  \& {Dolphin}}{{Murphy} et~al.}{2011}]{murphy2011}
{Murphy} J.~W.,  {Jennings} Z.~G.,  {Williams} B.,  {Dalcanton} J.~J.,
  {Dolphin} A.~E.,  2011, \mn@doi [\apjl] {10.1088/2041-8205/742/1/L4}, \href
  {http://adsabs.harvard.edu/abs/2011ApJ...742L...4M} {742, L4}

\bibitem[\protect\citeauthoryear{{Murphy}, {Khan}, {Williams}, {Dolphin},
  {Dalcanton}  \& {D{\'{\i}}az-Rodr{\'{\i}}guez}}{{Murphy}
  et~al.}{2018}]{murphy2018}
{Murphy} J.~W.,  {Khan} R.,  {Williams} B.,  {Dolphin} A.~E.,  {Dalcanton} J.,
   {D{\'{\i}}az-Rodr{\'{\i}}guez} M.,  2018, \mn@doi [\apj]
  {10.3847/1538-4357/aac2be}, \href
  {http://adsabs.harvard.edu/abs/2018ApJ...860..117M} {860, 117}

\bibitem[\protect\citeauthoryear{{Panagia}, {Romaniello}, {Scuderi}  \&
  {Kirshner}}{{Panagia} et~al.}{2000}]{panagia2000}
{Panagia} N.,  {Romaniello} M.,  {Scuderi} S.,   {Kirshner} R.~P.,  2000,
  \mn@doi [\apj] {10.1086/309212}, \href
  {https://ui.adsabs.harvard.edu/abs/2000ApJ...539..197P} {539, 197}

\bibitem[\protect\citeauthoryear{{Podsiadlowski}, {Langer}, {Poelarends},
  {Rappaport}, {Heger}  \& {Pfahl}}{{Podsiadlowski}
  et~al.}{2004}]{Podsiadlowski2004}
{Podsiadlowski} P.,  {Langer} N.,  {Poelarends} A.~J.~T.,  {Rappaport} S.,
  {Heger} A.,   {Pfahl} E.,  2004, \mn@doi [\apj] {10.1086/421713}, \href
  {https://ui.adsabs.harvard.edu/abs/2004ApJ...612.1044P} {612, 1044}

\bibitem[\protect\citeauthoryear{{Sarbadhicary}, {Badenes}, {Chomiuk},
  {Caprioli}  \& {Huizenga}}{{Sarbadhicary} et~al.}{2017}]{sarbadhicary2017}
{Sarbadhicary} S.~K.,  {Badenes} C.,  {Chomiuk} L.,  {Caprioli} D.,
  {Huizenga} D.,  2017, \mn@doi [\mnras] {10.1093/mnras/stw2566}, \href
  {http://adsabs.harvard.edu/abs/2017MNRAS.464.2326S} {464, 2326}

\bibitem[\protect\citeauthoryear{{Smartt}}{{Smartt}}{2009}]{smartt2009}
{Smartt} S.~J.,  2009, \mn@doi [\araa] {10.1146/annurev-astro-082708-101737},
  \href {http://adsabs.harvard.edu/abs/2009ARA%26A..47...63S} {47, 63}

\bibitem[\protect\citeauthoryear{{Smartt}}{{Smartt}}{2015}]{smartt2015}
{Smartt} S.~J.,  2015, \mn@doi [\pasa] {10.1017/pasa.2015.17}, \href
  {http://adsabs.harvard.edu/abs/2015PASA...32...16S} {32, e016}

\bibitem[\protect\citeauthoryear{{Smartt}, {Gilmore}, {Tout}  \&
  {Hodgkin}}{{Smartt} et~al.}{2002a}]{smartt2002em}
{Smartt} S.~J.,  {Gilmore} G.~F.,  {Tout} C.~A.,   {Hodgkin} S.~T.,  2002a,
  \mn@doi [\apj] {10.1086/324690}, \href
  {http://adsabs.harvard.edu/abs/2002ApJ...565.1089S} {565, 1089}

\bibitem[\protect\citeauthoryear{{Smartt}, {Vreeswijk}, {Ramirez-Ruiz},
  {Gilmore}, {Meikle}, {Ferguson}  \& {Knapen}}{{Smartt}
  et~al.}{2002b}]{smartt2002ap}
{Smartt} S.~J.,  {Vreeswijk} P.~M.,  {Ramirez-Ruiz} E.,  {Gilmore} G.~F.,
  {Meikle} W.~P.~S.,  {Ferguson} A.~M.~N.,   {Knapen} J.~H.,  2002b, \mn@doi
  [\apjl] {10.1086/341747}, \href
  {http://adsabs.harvard.edu/abs/2002ApJ...572L.147S} {572, L147}

\bibitem[\protect\citeauthoryear{{Smartt}, {Maund}, {Hendry}, {Tout},
  {Gilmore}, {Mattila}  \& {Benn}}{{Smartt} et~al.}{2004}]{Smartt2004}
{Smartt} S.~J.,  {Maund} J.~R.,  {Hendry} M.~A.,  {Tout} C.~A.,  {Gilmore}
  G.~F.,  {Mattila} S.,   {Benn} C.~R.,  2004, \mn@doi [Science]
  {10.1126/science.1092967}, \href
  {http://adsabs.harvard.edu/abs/2004Sci...303..499S} {303, 499}

\bibitem[\protect\citeauthoryear{{Smartt}, {Eldridge}, {Crockett}  \&
  {Maund}}{{Smartt} et~al.}{2009}]{smarttetal2009}
{Smartt} S.~J.,  {Eldridge} J.~J.,  {Crockett} R.~M.,   {Maund} J.~R.,  2009,
  \mn@doi [\mnras] {10.1111/j.1365-2966.2009.14506.x}, \href
  {http://adsabs.harvard.edu/abs/2009MNRAS.395.1409S} {395, 1409}

\bibitem[\protect\citeauthoryear{{Smith} et~al.,}{{Smith}
  et~al.}{2011}]{smith2011}
{Smith} N.,  et~al., 2011, \mn@doi [\apj] {10.1088/0004-637X/732/2/63}, \href
  {http://adsabs.harvard.edu/abs/2011ApJ...732...63S} {732, 63}

\bibitem[\protect\citeauthoryear{{Sukhbold}, {Ertl}, {Woosley}, {Brown}  \&
  {Janka}}{{Sukhbold} et~al.}{2016}]{sukhbold2016}
{Sukhbold} T.,  {Ertl} T.,  {Woosley} S.~E.,  {Brown} J.~M.,   {Janka} H.-T.,
  2016, \mn@doi [\apj] {10.3847/0004-637X/821/1/38}, \href
  {http://adsabs.harvard.edu/abs/2016ApJ...821...38S} {821, 38}

\bibitem[\protect\citeauthoryear{{Ugliano}, {Janka}, {Marek}  \&
  {Arcones}}{{Ugliano} et~al.}{2012}]{ugliano2012}
{Ugliano} M.,  {Janka} H.-T.,  {Marek} A.,   {Arcones} A.,  2012, \mn@doi
  [\apj] {10.1088/0004-637X/757/1/69}, \href
  {http://adsabs.harvard.edu/abs/2012ApJ...757...69U} {757, 69}

\bibitem[\protect\citeauthoryear{{Van Dyk}}{{Van Dyk}}{2017}]{vandyk2017}
{Van Dyk} S.~D.,  2017, \mn@doi [RSPTA] {10.1098/rsta.2016.0277}, \href
  {http://adsabs.harvard.edu/abs/2017RSPTA.37560277V} {375, 20160277}

\bibitem[\protect\citeauthoryear{{Van Dyk}, {Li}  \& {Filippenko}}{{Van Dyk}
  et~al.}{2003a}]{vandyk2003du}
{Van Dyk} S.~D.,  {Li} W.,   {Filippenko} A.~V.,  2003a, \mn@doi [\pasp]
  {10.1086/374299}, \href {http://adsabs.harvard.edu/abs/2003PASP..115..448V}
  {115, 448}

\bibitem[\protect\citeauthoryear{{Van Dyk}, {Li}  \& {Filippenko}}{{Van Dyk}
  et~al.}{2003b}]{vandyk2003gd}
{Van Dyk} S.~D.,  {Li} W.,   {Filippenko} A.~V.,  2003b, \mn@doi [\pasp]
  {10.1086/378308}, \href {http://adsabs.harvard.edu/abs/2003PASP..115.1289V}
  {115, 1289}

\bibitem[\protect\citeauthoryear{{Van Dyk} et~al.,}{{Van Dyk}
  et~al.}{2011}]{vandyk2011}
{Van Dyk} S.~D.,  et~al., 2011, \mn@doi [\apjl] {10.1088/2041-8205/741/2/L28},
  \href {http://adsabs.harvard.edu/abs/2011ApJ...741L..28V} {741, L28}

\bibitem[\protect\citeauthoryear{{Van Dyk} et~al.,}{{Van Dyk}
  et~al.}{2012a}]{vandyk2012a}
{Van Dyk} S.~D.,  et~al., 2012a, \mn@doi [\aj] {10.1088/0004-6256/143/1/19},
  \href {http://adsabs.harvard.edu/abs/2012AJ....143...19V} {143, 19}

\bibitem[\protect\citeauthoryear{{Van Dyk} et~al.,}{{Van Dyk}
  et~al.}{2012b}]{vandyk2012rsg}
{Van Dyk} S.~D.,  et~al., 2012b, \mn@doi [\apj] {10.1088/0004-637X/756/2/131},
  \href {http://adsabs.harvard.edu/abs/2012ApJ...756..131V} {756, 131}

\bibitem[\protect\citeauthoryear{{Vink{\'o}} et~al.,}{{Vink{\'o}}
  et~al.}{2009}]{vinko2009}
{Vink{\'o}} J.,  et~al., 2009, \mn@doi [\apj] {10.1088/0004-637X/695/1/619},
  \href {http://adsabs.harvard.edu/abs/2009ApJ...695..619V} {695, 619}

\bibitem[\protect\citeauthoryear{{Walmswell} \& {Eldridge}}{{Walmswell} \&
  {Eldridge}}{2012}]{Walmswell2012}
{Walmswell} J.~J.,  {Eldridge} J.~J.,  2012, \mn@doi [\mnras]
  {10.1111/j.1365-2966.2011.19860.x}, \href
  {https://ui.adsabs.harvard.edu/abs/2012MNRAS.419.2054W} {419, 2054}

\bibitem[\protect\citeauthoryear{{Wang}, {Yang}, {Zhang}, {Ma}, {Zhou}, {Li},
  {Lou}  \& {Li}}{{Wang} et~al.}{2005}]{wang2005}
{Wang} X.,  {Yang} Y.,  {Zhang} T.,  {Ma} J.,  {Zhou} X.,  {Li} W.,  {Lou}
  Y.-Q.,   {Li} Z.,  2005, \mn@doi [\apjl] {10.1086/431903}, \href
  {http://adsabs.harvard.edu/abs/2005ApJ...626L..89W} {626, L89}

\bibitem[\protect\citeauthoryear{{Williams}, {Peterson}, {Murphy}, {Gilbert},
  {Dalcanton}, {Dolphin}  \& {Jennings}}{{Williams}
  et~al.}{2014}]{williams2014}
{Williams} B.~F.,  {Peterson} S.,  {Murphy} J.,  {Gilbert} K.,  {Dalcanton}
  J.~J.,  {Dolphin} A.~E.,   {Jennings} Z.~G.,  2014, \mn@doi [\apj]
  {10.1088/0004-637X/791/2/105}, \href
  {http://adsabs.harvard.edu/abs/2014ApJ...791..105W} {791, 105}

\bibitem[\protect\citeauthoryear{Williams, Hillis, Murphy, Gilbert, Dalcanton
  \& Dolphin}{Williams et~al.}{2018}]{williams2018}
Williams B.~F.,  Hillis T.~J.,  Murphy J.~W.,  Gilbert K.,  Dalcanton J.~J.,
  Dolphin A.~E.,  2018, The Astrophysical Journal, 860, 39

\bibitem[\protect\citeauthoryear{{Williams}, {Hillis}, {Blair}, {Long},
  {Murphy}, {Dolphin}, {Khan}  \& {Dalcanton}}{{Williams}
  et~al.}{2019}]{williams2019}
{Williams} B.~F.,  {Hillis} T.~J.,  {Blair} W.~P.,  {Long} K.~S.,  {Murphy}
  J.~W.,  {Dolphin} A.,  {Khan} R.,   {Dalcanton} J.~J.,  2019, \mn@doi [\apj]
  {10.3847/1538-4357/ab2190}, \href
  {https://ui.adsabs.harvard.edu/abs/2019ApJ...881...54W} {881, 54}

\bibitem[\protect\citeauthoryear{{Woosley}, {Heger}  \& {Weaver}}{{Woosley}
  et~al.}{2002}]{woosley2002}
{Woosley} S.~E.,  {Heger} A.,   {Weaver} T.~A.,  2002, \mn@doi [Reviews of
  Modern Physics] {10.1103/RevModPhys.74.1015}, \href
  {http://adsabs.harvard.edu/abs/2002RvMP...74.1015W} {74, 1015}

\bibitem[\protect\citeauthoryear{{Zapartas} et~al.,}{{Zapartas}
  et~al.}{2017}]{zapartas2017}
{Zapartas} E.,  et~al., 2017, \mn@doi [\aap] {10.1051/0004-6361/201629685},
  \href {http://adsabs.harvard.edu/abs/2017A%26A...601A..29Z} {601, A29}

\bibitem[\protect\citeauthoryear{{Zapartas} et~al.,}{{Zapartas}
  et~al.}{2019}]{Zapartas2019}
{Zapartas} E.,  et~al., 2019, \mn@doi [\aap] {10.1051/0004-6361/201935854},
  \href {https://ui.adsabs.harvard.edu/abs/2019A&A...631A...5Z} {631, A5}

\bibitem[\protect\citeauthoryear{{Zapartas}, {de Mink}, {Justham}, {Smith},
  {Renzo}  \& {de Koter}}{{Zapartas} et~al.}{2020}]{Zapartas2020}
{Zapartas} E.,  {de Mink} S.~E.,  {Justham} S.,  {Smith} N.,  {Renzo} M.,   {de
  Koter} A.,  2020, arXiv e-prints, \href
  {https://ui.adsabs.harvard.edu/abs/2020arXiv200207230Z} {p. arXiv:2002.07230}

\bibitem[\protect\citeauthoryear{{de Mink}, {Sana}, {Langer}, {Izzard}  \&
  {Schneider}}{{de Mink} et~al.}{2014}]{demink2014merger}
{de Mink} S.~E.,  {Sana} H.,  {Langer} N.,  {Izzard} R.~G.,   {Schneider}
  F.~R.~N.,  2014, \mn@doi [\apj] {10.1088/0004-637X/782/1/7}, \href
  {https://ui.adsabs.harvard.edu/abs/2014ApJ...782....7D} {782, 7}

\makeatother
\end{thebibliography}

\label{lastpage}
\end{document}